\documentclass[sigconf, natbib=true]{acmart}
\usepackage{booktabs}
\usepackage{bookmark,xspace}
\usepackage{multirow}
\usepackage{paralist}
\usepackage{graphicx}
\usepackage{subfigure}
\usepackage{amsmath}
\usepackage{bbding}
\usepackage{enumitem}
\usepackage{color}
\usepackage{makecell}

\AtBeginDocument{%
  \providecommand\BibTeX{{%
    \normalfont B\kern-0.5em{\scshape i\kern-0.25em b}\kern-0.8em\TeX}}}

\newcommand{\name}{OpenSiteRec\xspace}

\setcopyright{acmcopyright}
\copyrightyear{2018}
\acmYear{2018}
\acmDOI{XXXXXXX.XXXXXXX}

\acmConference[Conference acronym 'XX]{Make sure to enter the correct
  conference title from your rights confirmation emai}{June 03--05,
  2018}{Woodstock, NY}
\acmPrice{15.00}
\acmISBN{978-1-4503-XXXX-X/18/06}

\begin{document}

\title{OpenSiteRec: An Open Dataset for Site Recommendation}

\author{Xinhang Li}
\authornote{This work was done during the visiting at Nanyang Technological University when the author studied at Tsinghua University.}
\affiliation{%
 \institution{Tsinghua Univerisity}
 \city{Beijing}
 \country{China}
}
\email{xh-li20@mails.tsinghua.edu.cn}

\author{Xiangyu Zhao}
\affiliation{%
 \institution{City Univerisity of Hong Kong}
 \country{Hong Kong}
}
\email{xianzhao@cityu.edu.hk}

\author{Yejing Wang}
\affiliation{%
 \institution{City Univerisity of Hong Kong}
 \country{Hong Kong}
}
\email{yejing.wang@my.cityu.edu.hk}

\author{Yu Liu}
\affiliation{
 \institution{Tsinghua University}
 \city{Beijing}
 \country{China}
}
\email{liuyu2419@126.com}

\author{Yong Li}
\affiliation{
 \institution{Tsinghua University}
 \city{Beijing}
 \country{China}
}
\email{liyong07@tsinghua.edu.cn}

\author{Cheng Long}
\authornote{Cheng Long and Yong Zhang are the corresponding authors.}
\affiliation{%
 \institution{Nanyang Technological University}
 \country{Singapore}
}
\email{c.long@ntu.edu.sg}

\author{Yong Zhang}
\authornotemark[2]
\affiliation{%
 \institution{Tsinghua Univerisity}
 \city{Beijing}
 \country{China}
}
\email{zhangyong05@tsinghua.edu.cn}

\author{Chunxiao Xing}
\affiliation{%
 \institution{Tsinghua Univerisity}
 \city{Beijing}
 \country{China}
}
\email{xingcx@tsinghua.edu.cn}

\renewcommand{\shortauthors}{Xinhang Li et al.}

\begin{abstract}
As a representative information retrieval task, site recommendation, which aims at predicting the optimal sites for a brand or an institution to open new branches in an automatic data-driven way, is beneficial and crucial for brand development in modern business.
However, there is no publicly available dataset so far and most existing approaches are limited to an extremely small scope of brands, which seriously hinders the research on site recommendation.
Therefore, we collect, construct and release an open comprehensive dataset, namely \name, to facilitate and promote the research on site recommendation.
Specifically, \name leverages a heterogeneous graph schema to represent various types of real-world entities and relations in four international metropolises.
To evaluate the performance of the existing general methods on the site recommendation task, we conduct benchmarking experiments of several representative recommendation models on \name.
Furthermore, we also highlight the potential application directions to demonstrate the wide applicability of \name.
We believe that our \name dataset is significant and anticipated to encourage the development of advanced methods for site recommendation.
\name is available online at https://OpenSiteRec.github.io/.
\end{abstract}

\ccsdesc[500]{Information systems~Data mining}
\ccsdesc[500]{Information systems~Business intelligence}
\ccsdesc[500]{Information systems~Learning to rank}
\ccsdesc[500]{Information systems~Recommender systems}
\ccsdesc[500]{Information systems~Location based services}

\keywords{site recommendation, dataset, heterogeneous graph, benchmark}
\maketitle
\begin{table*}
    \caption{Comparison with the datasets used by other site recommendation approaches. Specifically, `/' between numbers denotes the split w.r.t. different brands, `*' denotes the approximate number w.r.t. the setting in the corresponding paper. While the other approaches only focus on a small scope of brands, \name covers a much larger number of brands with more comprehensive information.}
    \centering
    \begin{tabular}{lcc|c|rrr}
        \toprule
        \textbf{Approach} & \textbf{Venue} & \textbf{Year} & \textbf{City} & \textbf{Brand} & \textbf{Region} & \textbf{Site}\\
        \midrule
        Geo-Spotting~\cite{DBLP:conf/kdd/KaramshukNSNM13} & KDD & 2013 & New York City & 3 & 32* & 186/104/66 \\
        \midrule
        \multirow{2}*{ANNRR~\cite{DBLP:conf/huc/ChenZPMYKZL15}} & \multirow{2}*{UbiComp} & \multirow{2}*{2015} & Washington, D.C. & 1 & 181 & 203 \\
        & & & Hangzhou & 1 & 882 & 2,115\\
        \midrule
        PAM~\cite{DBLP:conf/gis/LiZJWUG15} & SIGSPATIAL & 2015 & Tianjin & 1 & 99,007 & 34\\
        \midrule
        BL-G-CoSVD~\cite{DBLP:journals/tkdd/YuTWGM16} & TKDD & 2016 & Shanghai & 5 & 17,435 & 17,435\\
        \midrule
        DD3S~\cite{DBLP:conf/gis/XuWWZLW16} & SIGSPATIAL & 2016 & Beijing & 2/2 & 10* & 1,882/1,343\\
        \midrule
        \multirow{4}*{CityTransfer~\cite{DBLP:journals/imwut/GuoLZWY17}} & \multirow{4}*{UbiComp} & \multirow{4}*{2017} & Beijing & \multirow{4}*{3} & \multirow{4}*{1,000-3,000*} & 123/160/179\\
        & & & Shanghai & & & 147/46/156\\
        & & & Xi'an & & & 69/59/189\\
        & & & Nanjing & & & 73/46/57\\
        \midrule
        DeepStore~\cite{DBLP:journals/iotj/LiuGLZCZLYZY19} & IOT & 2019 & 13 Cities & 49 in Total & 300* & 49 in Total\\
        \midrule
        WANT~\cite{DBLP:journals/tkdd/LiuGZZCHZZY21} & TKDD & 2021 & 6 Cities & 100* in Total & 300* & 100* in Total\\
        \midrule
        \multirow{2}*{UrbanKG~\cite{DBLP:journals/corr/abs-2111-00787}} & \multirow{2}*{ArXiv} & \multirow{2}*{2021} & Beijing & 398 & 528 & 22,468 \\
        & & & Shanghai & 441 & 2,042 & 38,394 \\
        \midrule
        $O^2$-SiteRec~\cite{DBLP:conf/icde/YanWYGHZ22} & ICDE & 2022 & Shanghai & 122 & 2,000* & 39,465\\
        \midrule
        \multirow{2}*{UUKG~\cite{DBLP:journals/corr/abs-2306-11443}} & \multirow{2}*{ArXiv} & \multirow{2}*{2023} & New York City & 15 & 260 & 62,450 \\
        & & & Chicago & 15 & 77 & 31,573 \\
        \midrule
        \multirow{4}*{\textbf{\name}} & \multirow{4}*{/} & \multirow{4}*{/} & Chicago & 969 & 801 & 8,044\\
        & & & New York City & 2,702 & 2,325 & 14,189 \\
        & & & Singapore & 1,922 & 2,043 & 9,912\\
        & & & Tokyo & 4,861 & 3,036 & 26,765\\
        \bottomrule
    \end{tabular}
    \label{tab:comparison}
\end{table*}

\section{Introduction}

In modern business, selecting an optimal site to open a new branch is definitely crucial for the development of a brand or an institution~\cite{Huff1966APS,Hernndez2000TheAA,ahin2019AnalyticHP}.
An appropriate site will bring substantial profits while an inappropriate site may lead to business failure~\cite{Kumar2000TheEO,Wu2016OptimalSS,Wu2017ADF}.
Thus, properly determining the best choice from so many candidate sites is quite important yet complex, since it needs to take many factors into accounts~\cite{McFadden1977ModellingTC,Marsh1994EquityMI,Shao2020ARO}, such as the brand types and the population surrounding the site.
Typically, this task is mainly accomplished by the professional consulting or marketing departments of companies~\cite{Revelle2005LocationAA}, which is usually labor-intensive and time-consuming.
Meanwhile, human error and bias can also lead to suboptimal solutions.
Therefore, it is difficult for such an artificial approach to fulfill the high demand of rapid development in modern business.

Thanks to the booming development in information retrieval, automatic data-driven approaches have been introduced to assist the decision-making and reduce the cost, i.e. site recommendation~\cite{Church2008BusinessSS,DBLP:conf/kdd/KaramshukNSNM13,Liu2022DevelopingKG}.
These approaches come with a wide variety of definitions of site recommendation, including the association analysis~\cite{DBLP:conf/kdd/KaramshukNSNM13} for feature selection, the rating prediction~\cite{DBLP:journals/imwut/GuoLZWY17}, the consumption prediction~\cite{DBLP:journals/iotj/LiuGLZCZLYZY19,DBLP:journals/tkdd/LiuGZZCHZZY21} and the top-N recommendation~\cite{DBLP:conf/gis/LiZJWUG15,DBLP:journals/tkdd/YuTWGM16,DBLP:journals/corr/abs-2111-00787}.
While they share the idea of treating the site recommendation problem as a ranking task, their significantly different definitions make it difficult for them to be compared directly.
Thus, all these works are independent of each other and they have to undesirably start from scratch instead of making continuous improvements, which is detrimental to the subsequent research on site recommendation.
Meanwhile, their datasets only cover very small yet different scopes in site recommendation, such as bike sharing station~\cite{DBLP:conf/huc/ChenZPMYKZL15}, chain hotel~\cite{DBLP:journals/imwut/GuoLZWY17} and online stores with courier capacity~\cite{DBLP:conf/icde/YanWYGHZ22}.
This leads to failure in utilizing comprehensive information across scopes and severe data sparsity problems in site recommendation.
Even more unfortunately, none of them has released their datasets so far.
In most cases, collecting data and creating dataset are necessary in research but low-yielding since the dataset is typically a fundamental part in a scientific thesis.
Therefore, the lack of publicly available dataset brings inconvenience to the researchers and forces them to spend long time dealing with the dataset construction, which even hinders the development of site recommendation solutions.
According to the above problems, we believe a unified definition and a comprehensive open dataset of site recommendation are necessary and crucial for the benign development of the following research on site recommendation.

To this end, we propose a formal problem definition of site recommendation by jointly considering and summarizing the definitions of existing studies.
Based on this problem definition, we collect, construct and release an \textbf{Open} benchmarking dataset for \textbf{Site} \textbf{Rec}ommendation, namely \textbf{\name}.
Specifically, \name consists of four international metropolises, including \textit{Chicago}, \textit{New York City}, \textit{Singapore} and \textit{Tokyo}.
Different from the datasets used by the existing approaches, our proposed \name contains all the brands and regions from all the scopes and types in the whole cities and thus yields a wide-range, much larger and more comprehensive dataset.
Meanwhile, \name provides sufficient trustworthy commercial relationships and organizes the different types of real-world concepts into a heterogeneous graph to offer more comprehensive information.
Furthermore, we also conduct benchmarking experiments of several representative baselines on \name to facilitate future research.
Some discussions of the potential application directions of \name in other research areas, including brand entry forecasting and business area planning, and also the limitations of \name are presented to give a broader view of \name.

The contributions of this paper are summarized as follows:
\begin{itemize}[leftmargin=*]
    \item We introduce a formal definition of site recommendation by summarizing the task definitions of existing works, which unifies them to provide open benchmarks for the following research.
    \item We collect, construct and release an open comprehensive dataset of four international metropolises, namely \textbf{\name}, to facilitate the subsequent research on site recommendation. To the best of our knowledge, \name is the first publicly available dataset for site recommendation.
    \item We conduct benchmarking experiments of 16 widely-used baseline models in recommendation on \name, to verify their effectiveness in site recommendation and to facilitate future research for comparison.
    \item Besides site recommendation, various other research areas such as brand expansion, urban planning and facility location can benefit from \name given that it embeds rich information on both commercial and geographical aspects of urban spaces.
\end{itemize}

\begin{figure*}
    \centering
    \includegraphics[width=\linewidth]{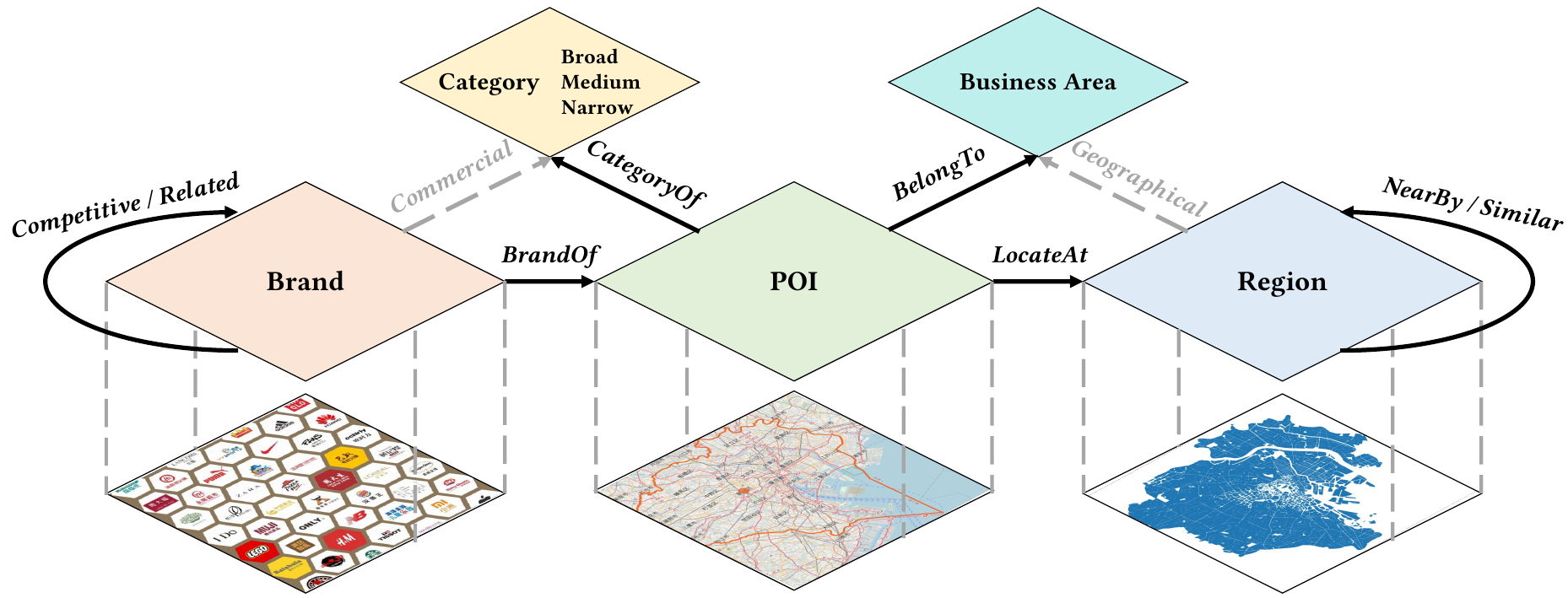}
    \caption{Schema of OpenSiteRec. The different types of entities and relations are represented by the polygons and arrows respectively. The solid arrows denote the definite relations between different entity types while the dotted arrows denote the indefinite relations that are not exactly defined but can be derived from existing information.}
    \label{fig:schema}
\end{figure*}

\section{Related Works}

Site recommendation for store brands and public facilities is a widely-studied problem with strong practical significance in modern business~\cite{DBLP:journals/cor/BermanK02} and urban planning~\cite{Rathore2016UrbanPA}.
The most seminal attempt~\cite{Jensen2006NetworkbasedPO,DBLP:conf/ida/Jensen09} proposes to investigate the potential effects of different features for retail store site identification.
Following it, some data mining approaches are proposed in evaluating the correlations between the street centrality and the geographical distribution of activities in Bologna~\cite{Porta2009StreetCA} and Barcelona~\cite{Porta2012StreetCA}.

With the booming development of machine learning algorithms, applying them for site recommendation becomes an effective and efficient solution, which has attracted increasing interest and thus yields many approaches.
Unfortunately, not only their datasets are limited to small scopes but also none of them has released their datasets.
The detailed comparison between the datasets of these approaches and our \name is shown in Table~\ref{tab:comparison}.

Geo-Spotting~\cite{DBLP:conf/kdd/KaramshukNSNM13} first extends the early data mining approaches with machine learning algorithms to better analyze the effectiveness of geographical and mobility features in site recommendation.
This approach focuses on 3 fast food brands in New York City and analyzes the key factors of site recommendation separately for each brand with sufficient samples.
ANNRR~\cite{DBLP:conf/huc/ChenZPMYKZL15} proposes a semi-supervised feature selection method on heterogeneous urban data to predict bike trip demand for the bike sharing station recommendation in two cities.
In this work, the site recommendation is limited to bike sharing station.
PAM~\cite{DBLP:conf/gis/LiZJWUG15} utilizes the traffic information with partitioning around medoids algorithm to determine the optimal location of new ambulance station.
Although there are a great number of candidate regions, there are only 34 existing stations available for training that may do harm to the credibility of results.
BL-G-CoSVD~\cite{DBLP:journals/tkdd/YuTWGM16} introduces bias learning and integrates both location and commercial features into SVD to recommend the suitable shop-type for each site.
Contrary to the other approaches that predict the optimal region for a given brand, this approach predicts the shop-type for a given region with only 5 candidate shop-types so that the difficulty of this task is much lower.
DD3S~\cite{DBLP:conf/gis/XuWWZLW16} learns to rank the candidate demand centers for two coffee shop brands and two chain hotel brands by predicting the number of customers at the given location with multiple spatial-temporal data sources.
Since the candidate demand centers are predetermined by clustering the demand and supply gaps, there are only about 10 regions being determined as candidates, which is a small number.
DeepStore~\cite{DBLP:journals/iotj/LiuGLZCZLYZY19} leverages a deep neural network on both dense and sparse features for predicting the consumption level of 49 stores in their surrounding areas of 13 cities to recommend the optimal site.
However, the limited amount of data significantly increases the dependence of this task to the quality of features. 

While these approaches consider the site recommendation of different cities individually, there are also some works focusing on knowledge transfer across different cities. 
CityTransfer~\cite{DBLP:journals/imwut/GuoLZWY17} transfers knowledge from a source city to a target city via both inter- and intra-city views for site recommendation of 3 chain hotel brands in a new city.
Specifically, it chooses Beijing or Shanghai as the source city and Xi'an or Nanjing as the target city to improve the performance on the cities with less stores.
WANT~\cite{DBLP:journals/tkdd/LiuGZZCHZZY21} employs adversarial learning to diminish the distribution discrepancy between the source city and target city for predicting the consumption of stores in given areas.
Then, it ranks the candidate areas in the target city according to the consumption for site recommendation.
Similar to DeepStore, the dataset used by WANT is also small in size.

Despite their success, all these traditional site recommendation approaches rely heavily on feature engineering with simple model structures, especially the fine-grained manually-crafted features~\cite{DBLP:conf/kdd/KaramshukNSNM13,DBLP:journals/tkdd/YuTWGM16,DBLP:conf/gis/XuWWZLW16,DBLP:journals/imwut/GuoLZWY17,DBLP:journals/iotj/LiuGLZCZLYZY19,DBLP:journals/tkdd/LiuGZZCHZZY21}, which are hard to design and may introduce human biases.
Different from them, the recent approaches pay efforts in utilizing complex models to automatically capture the latent features from multi-source data for site recommendation, which is a new research trend of site recommendation.
UrbanKG~\cite{DBLP:journals/corr/abs-2111-00787} constructs a knowledge graph from urban data, built upon which a relational graph neural network model is designed for efficient and effective site recommendation.
UrbanKG utilizes a comprehensive dataset and the well-defined data structure is suitable for the research of site recommendation.
However, this dataset focuses on the giant brands and has not been released for publicly available.
$O^2$-SiteRec~\cite{DBLP:conf/icde/YanWYGHZ22} conducts site recommendation by ranking the candidate regions with order number and delivery time from the courier capacity perspective in Online-to-Offline (O2O) stores of delivery platforms.
This approach only uses the data of a small scope of brands from a single city, which is not sufficient.
UUKG~\cite{DBLP:journals/corr/abs-2306-11443} proposes an urban knowledge graph as the foundation for downstream spatial-temporal prediction tasks.
Although UUKG is not designed for site recommendation, the link prediction task on it shares many similarities with site recommendation.
Differently, it only have 15 broad categories to distinguish the POIs rather than  fine-grained brands and more than 80\% of the POIs are from merely 2 categories.
Therefore, the vanilla UUKG is not sufficient to well support site recommendation.

Compared with their datasets, our proposed \name has outstanding superiorities from four perspectives:
\begin{itemize}[leftmargin=*]
    \item \textbf{Wide-range and Large-scale}: Existing site recommendation approaches mainly consider a small range of brands or focus on a specific scenario, such as chain hotel and courier capacity. Therefore, they usually collect data only for the specific demand, which yields a small size of data in the final task. In contrast, \name takes all the brands and regions in the city into account and thus yields much larger data size.
    \item \textbf{Rich Commercial Features}: Existing approaches focus more on leveraging the features of regions for prediction while paying less attention to the features of brands. In contrast, \name provides more commercial features of brands and institutions via various types of relations among them.
    \item \textbf{Comprehensive Information}: Most of the existing approaches manually define the fine-grained features using original geographical and demographical information, which may lead to human bias and information loss. In contrast, \name models the data as a heterogeneous graph with different types of nodes and edges, which provides more comprehensive information, such as the competitive relations between brands, along with the original features.
    \item \textbf{Publicly Available}: None of the existing studies have released their datasets. To the best of our knowledge, \name is the first publicly available dataset for site recommendation. The released \name dataset will encourage continuous research in site recommendation.
\end{itemize}

\section{Data Description}
In this section, we will first introduce the schema definition to better illustrate the overview of \name. Then, we will describe the whole process of data construction in detail. Finally, we will provide the statistics along with the usage of \name.

\subsection{Problem Definition}

While most of the existing works formulate the site recommendation as a ranking task ultimately, their data is usually used for a specific purpose and thus it yields different definitions of site recommendation, such as store-type recommendation~\cite{DBLP:journals/tkdd/YuTWGM16}, consumption level prediction~\cite{DBLP:journals/iotj/LiuGLZCZLYZY19} and knowledge graph link prediction~\cite{DBLP:journals/corr/abs-2111-00787}.
From the higher level of perspective in abstraction, all these existing definitions of site recommendation share the same prototype.
No matter what kinds of aspects of information they use or forms of task they apply, the ultimate objective of them is the same to obtain the ranking list of sites.
Since there is no precise consentient definition of site recommendation, we propose a formal definition for site recommendation task by considering the common parts among all the definitions of existing works as follows.

\begin{definition}
    Let $\mathcal{B} = \{b_1, b_2, ..., b_M\}$ denote the brand set with $M$ brands and $\mathcal{R} = \{r_1, r_2, ..., r_N\}$ denote the region set with $N$ regions.
    Each POI that belongs to brand $b_i$ and locates at region $r_j$ contributes to a value $P_{ij} = 1$ in the matrix $\mathbf{P} = \{0, 1\} \in \mathbb{R}^{M \times N}$.
    Therefore, the site recommendation task aims at predicting a ranking list of candidate regions for each given brand.
\end{definition}

Different from the definitions of other studies, our definition is more general that unifies the different concepts, such as brand, store and shop type, and requires no additional definition, e.g. the relation in knowledge graph link prediction.
Moreover, such a definition is more straightforward that directly considers the ultimate ranking objective rather than converting the other objectives to ranking.

\begin{table*}
    \centering
    \caption{Annotation of relation definitions in \name.}
    \begin{tabular}{c|c|c|c}
        \toprule
        \textbf{Relation} & \textbf{Subject \& Object Ontology} & \textbf{Symmetry} & \textbf{Data Source}\\
        \midrule
        BrandOf & (Brand, POI) & \XSolidBrush & Commercial Data, Site Data\\
        LocateAt & (POI, Region) & \XSolidBrush & Geographical Data, Site Data\\
        CategoryOf & (POI, Category\_) & \XSolidBrush & Commercial Data, Site Data\\
        BelongTo & (POI, Business Area) & \XSolidBrush & Geographical Data, Site Data\\
        Competitive & (Brand, Brand) & \Checkmark & Commercial Data\\
        Related & (Brand, Brand) & \Checkmark & Commercial Data\\
        NearBy & (Region, Region) & \Checkmark & Geographical Data\\
        Similar & (Region, Region) & \Checkmark & Geographical Data\\
        SubCategoryOf & (Category\_, Category\_) & \XSolidBrush & Commercial Data\\
        \bottomrule
    \end{tabular}
    \label{tab:relations}
\end{table*}

\subsection{Schema Definition}

In order to clarify the goal of data to collect and the final structure of dataset to construct, we deliver a schema as shown in Figure~\ref{fig:schema} for illustration.
Such an overall schema has a graph structure, which consists of different types of entities to represent the real-world concepts and different types of edges between entities to indicate the commercial or geographical relations.
Since there is much different domain-specific information for each kind of brands or scenarios that provides additional complexity as mentioned above, such as the house price in site recommendation for chain hotel brands~\cite{DBLP:journals/imwut/GuoLZWY17}, we only consider the general and fundamental information in \name according to our proposed problem definition.
Based on this schema, we elaborate the definition of each kind of entities and relations in details in the following paragraphs.

To be in line with the aforementioned problem and schema definition, we first define five types of entities to denote the real-world concepts following UrbanKG~\cite{DBLP:journals/corr/abs-2111-00787}:
\begin{itemize}[leftmargin=*]
    \item \textit{Brand}: Brands denote either commercial brands in business that own multiple branches, e.g., Starbucks and Apple, or institutions that refer to special functions, e.g., Columbia University and United States Postal Service.
    \item \textit{Category}: Categories represent the functions of the venues. Due to the significant functional differences between venues, we define three levels of categories (broad, medium and narrow) for classification. For example, a Starbucks store has the categories of `Food and Beverage', `Beverage Shop' and `Coffee and Tea Shop' for broad, medium and narrow respectively.
    \item \textit{POI}: POIs are the basic functional venues in a city, such as shops, restaurants and schools. Each POI has sufficient commercial and geographical information.
    \item \textit{Business Area}: Business areas denote the special planned areas for business to form the scale effect, where the venues are usually very dense.
    \item \textit{Region}: Regions refer to the geographical divisions planned by the city governments. The principle and granularity of region division varies depending on the city governments. For example, the regions in Chicago are relatively large with only numbers for reference while the regions in Tokyo are much smaller with specific names like `Ginza 1 Chōme'.
\end{itemize}

Specifically, these five types of entities can be further categorized into three aspects by the types of their data source, which are commercial data (\textit{Brand} and \textit{Category}), site data (\textit{POI}) and geographical data (\textit{Business Area} and \textit{Region}).
From this perspective, we define the relation types between these five types of entities as Table~\ref{tab:relations}.
Each \textit{POI} can be mapped to a \textit{Brand} and a \textit{Region} from the commercial and geographical aspects of it.
Therefore, the \textit{POI} serves as a bridge to connect \textit{Brand} and \textit{Region} to construct the dataset for site recommendation.
Meanwhile, the \textit{Category} and \textit{Business Area} of a \textit{POI} are uniquely determined.
However, according to the real-world situation, a \textit{Brand} may have multiple relations of \textit{Category} and a \textit{Business Area} may consist of several parts from multiple instances of \textit{Region}.
So the relations between \textit{Brand} and \textit{Category}, \textit{Region} and \textit{Business Area} can not be exactly defined.

Moreover, there are also plenty of relations within \textit{Brand} (\textit{Competitive} and \textit{Related}) and relations within \textit{Region} (\textit{NearBy} and \textit{Similar}) that may be useful for site recommendation as follows:
\begin{itemize}[leftmargin=*]
    \item \textit{Competitive}: This commercial relation means that two different brands are competitive, e.g., KFC and McDonald's.
    \item \textit{Related}: This commercial relation means that two different brands belong to the same company or group, e.g., KFC and Pizza Hut both belong to Yum! Brands.
    \item \textit{NearBy}: This geographical relation means that two different regions are close in physical space, e.g., `Ginza 1 Chōme' and `Ginza 2 Chōme'.
    \item \textit{Similar}: This commercial relation means that two different regions have similar distributions of POI category, e.g., `Ginza 1 Chōme' and `Toranomon 4 Chōme'.
\end{itemize}

\begin{table*}
    \caption{Statistics of OpenSiteRec.}
    \centering
    \begin{tabular}{l|rrrrrrrrr}
        \toprule
        \multirow{2}*{\textbf{City}} & \multirow{2}*{\textbf{\makecell[r]{Geographic\\Unit}}} &
        \multirow{2}*{\textbf{POI}} & \multirow{2}*{\textbf{\makecell[r]{Business\\Area}}} &
        \multirow{2}*{\textbf{Region}} &
        \multirow{2}*{\textbf{\makecell[r]{Site\\(POI with Brand)}}} & 
        \multirow{2}*{\textbf{Brand}} & 
        \multicolumn{3}{c}{\textbf{Category}} \\
        \cmidrule{8-10}
        & & & & & & & \textbf{Broad} & \textbf{Medium} & \textbf{Narrow} \\
        \midrule
        Chicago & 1,471,416 & 16,154 & 77 & 801 & 8,044 & 969 & 10 & 39 & 107\\
        New York City & 2,678,932 & 41,403 & 71 & 2,325 & 14,189 & 2,702 & 10 & 43 & 129\\
        Singapore & 592,663 & 18,580 & 387 & 2,043 & 9,912 & 1,922 & 10 & 39 & 116\\
        Tokyo & 1,427,914 & 60,042 & / & 3,036 & 26,765 & 4,861 & 10 & 37 & 98\\
        \bottomrule
    \end{tabular}
    \label{tab:dataset}
\end{table*}

\subsection{Data Construction}

Due to the requirements of data quality to produce reliable dataset, we choose four international metropolises for \name at present considering their information comprehensiveness and data integrity, which are \textit{Chicago}, \textit{New York City}, \textit{Singapore} and \textit{Tokyo}.
In order to better support and promote the research for site recommendation, we plan to add more cities into \name on the premise of ensuring data quality in the future.
Note that all of the data is collected from open-source data sources and our \name is distributed under the same licence with them to fulfill the ethical regulations. 

On the basis of the schema, we collect data separately from different sources for the aforementioned three aspects.
First, the site data, which is the core to connect commercial data and geographical data, is obtained by extracting the POIs from OpenStreetMap\footnote{https://www.openstreetmap.org/}, licensed under the Open Data Commons Open Database License (ODbL)~\footnote{https://opendatacommons.org/licenses/odbl/}.
OpenStreetMap~\cite{DBLP:journals/pervasive/HaklayW08} is an open-source community-built map service that consists of three types of geographic unit, including nodes, ways and relations.
Typically, each node denotes a geographical point in a map while each way and each relation represents a line and a polygon that consist of series of points.
Specifically, we extract the data from the data distribution service\footnote{https://download.bbbike.org/osm/bbbike/} on December 1st, 2022.
For each geographic unit, OpenStreetMap provides a series of tags to describe its characteristics, such as name, brand and amenity.
Here we filter the POIs by extracting all the geographic units with at least one type of name tag since the nameless objects are mostly meaningless either.
To represent the geographical location of each POI to identify its region, we convert the original geographical information of different objects into their centroids' coordinates, which are pairs of longitude and latitude.

Then, we obtain the commercial data by assigning the POIs to the brands from Wikidata\footnote{https://www.wikidata.org/}.
For the POIs that already have the brand tags, it is as simple as directly taking the information from the tags like the wikidata code to extract their brands.
Unfortunately, the majority of POIs don't have any brand tags and the brand information is mainly contained in their names.
Meanwhile, there may be multiple brand names that correspond to the same brand.
Therefore, it is essential to design an effective method for brand matching.
In order to achieve more accurate and reliable brand matching, we apply a combination of phonetic matching and text matching algorithms.
Specifically, we utilize the soundex~\cite{DBLP:conf/sigir/ZobelD96} algorithm for phonetic matching and the Jaro distance~\cite{Jaro1989AdvancesIR,DBLP:conf/ijcai/CohenRF03} as edit distance for text matching.
For each brand name, we first translate it to Unicode by its pronunciation, which means the Japanese words are translated using Hepburn romanization and the Chinese words are translated using Hanyu Pinyin.
Two brand names are defined as matching only when they are matched by both two algorithms.
Here, we implement the matching algorithms using the open-source library jellyfish~\footnote{https://pypi.org/project/jellyfish/} and apply Jaro distance with $0.8$ as threshold of matching.
After the brand matching, the brands are grouped together into several disjoint sets according to their matches and each set will choose one specific brand name which has record in Wikidata as the corresponding name for the whole set.
Through this process, we can successfully obtain the precise brand of POIs without ambiguity.
For the sake of ensuring the reliability of such an automatic brand matching approach, we manually evaluate 5\% of the matched brands in each city and the evaluation results are quite convincing.
For each POI, the hierarchical categories are determined based on its functional tags and the brand information from Wikidata.
Specifically, the broad category and medium category types are manually defined and the narrow category types are directly extracted from Wikidata.
Take a branch store of Starbucks as example, the broad category `Food and Beverage' is determined by its functional tags `amenity=restaurant' and the narrow category `Cafe and Tea Shop' is obtained from the brand information of Starbucks in Wikidata while the medium category is then defined to be `Beverage Shop'.

Next, the geographical data is collected from the free public data on the data portal of Chicago\footnote{https://data.cityofchicago.org/}, New York City\footnote{https://opendata.cityofnewyork.us/}, Singapore\footnote{https://data.gov.sg/} and Tokyo\footnote{https://www.data.go.jp/} governments.
Specifically, the Tokyo government has not released the business area planning data so we omit the business area data of Tokyo.
Typically, the data is officially released by the government agencies of each city and thus the region partitions and the business area plannings are credible.
For the officially defined regions and business areas, we formulate them into polygon boundaries which consist of a series of coordinates.
Subsequently, we determine the geographical assignment of POIs by computing the inclusion relationships between the boundaries of the regions or business areas and the centroids of the POIs.

After obtaining all the entities, we further identify the intra-relations within \textit{Brand} and \textit{Region} to provide more comprehensive commercial and geographical information.
For the relations within \textit{Brand}, i.e. \textit{Competitive} and \textit{Related}, we directly extract them by the provided statements from Wikidata.
Specifically, we define all the other brands in the same lowest category as \textit{Competitive} and define the brands in `See Also' section as \textit{Related}.
Since these two relations are obtained from Wikidata, it is not abnormal for a brand to have no provided \textit{Competitive} or \textit{Related} brand.
For the relations within \textit{Region}, we calculate the shortest distances between pairs of regions to identify \textit{NearBy} relations with a maximum threshold $0.5$km and we calculate the cosine similarity of narrow category distributions of POIs to identify \textit{Similar} relations with a minimum threshold $0.9$.
Let the POI category distribution of region $i$ be $dist_i = [n_1, n_2, ..., n_K]$ where $n_k$ is the POI number of category $k$ in region $i$ and $K$ is the number of narrow category types, the cosine similarity between region $i$ and $j$ is defined as $sim=\frac{dist_i \cdot dist_j}{|dist_i||dist_j|}$.
Also, it is common for a region to have no \textit{Similar} region.

\subsection{Statistics and Usage}

\begin{table*}[h]
    \caption{An example of an instance in OpenSiteRec.}
    \centering
    \begin{tabular}{cccc}
        \toprule
        \multirow{2}*{\textbf{General}} & \textbf{ID} & \textbf{Name} & \textbf{City}\\
        & 259964052 & Starbucks	& Tokyo\\
        \midrule
        \multirow{4}*{\textbf{Commercial}} 
        & \textbf{Cate 1} & \textbf{Cate 2} & \textbf{Cate 3} \\
        & Food and Beverage & Beverage Shop & Coffee and Tea Shop\\
        \cmidrule{2-4}
        & \textbf{Brand} & \textbf{Competitive} & \textbf{Related} \\
        & Starbucks & Doutor, Tully's Coffee, ... & McDonald's, KFC, ...\\
        \midrule
        \multirow{4}*{\textbf{Geographical}} 
        & \textbf{Longitude} & \textbf{Latitude} & \textbf{District} \\
        & 139.7390476 & 35.684236 & Chiyoda City\\
        \cmidrule{2-4}
        & \textbf{Region} & \textbf{NearBy} & \textbf{Similar}\\
        & Kōjimachi 3 Chōme & Kōjimachi 6 Chōme, Kioichō, ... & Kōjimachi 4 Chōme, ...\\
        \bottomrule
    \end{tabular}
    \label{tab:example}
\end{table*}

\begin{figure*}[h]
    \centering
    \includegraphics[width=0.7\linewidth]{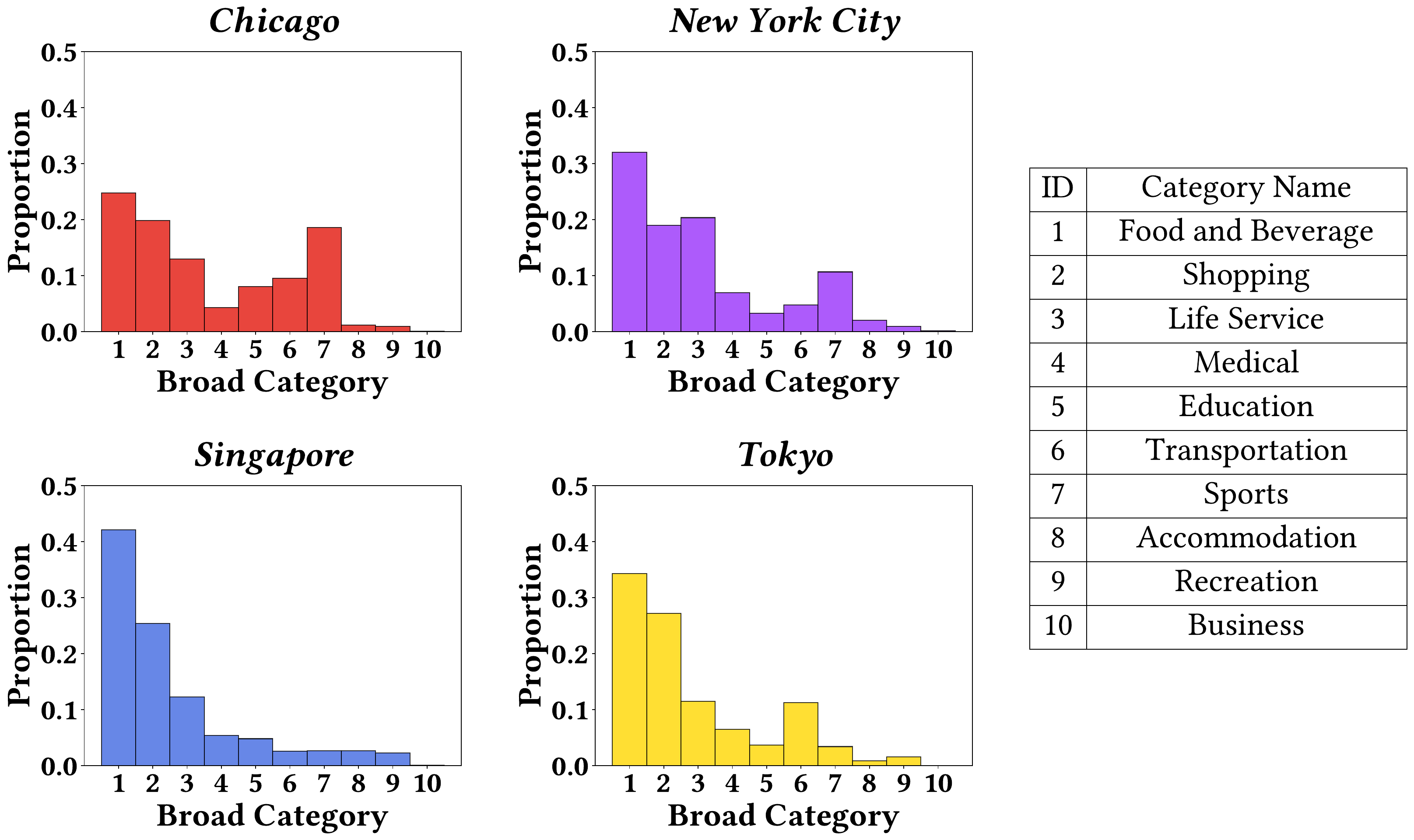}
    \caption{Distributions of site counts of categories in different cities. There are slight differences on the distributions among these four cities, which indicates the differences in urban planning and lifestyles of people.}
    \label{fig:dist_category}
\end{figure*}

\begin{figure}[h]
    \centering
    \includegraphics[width=\linewidth]{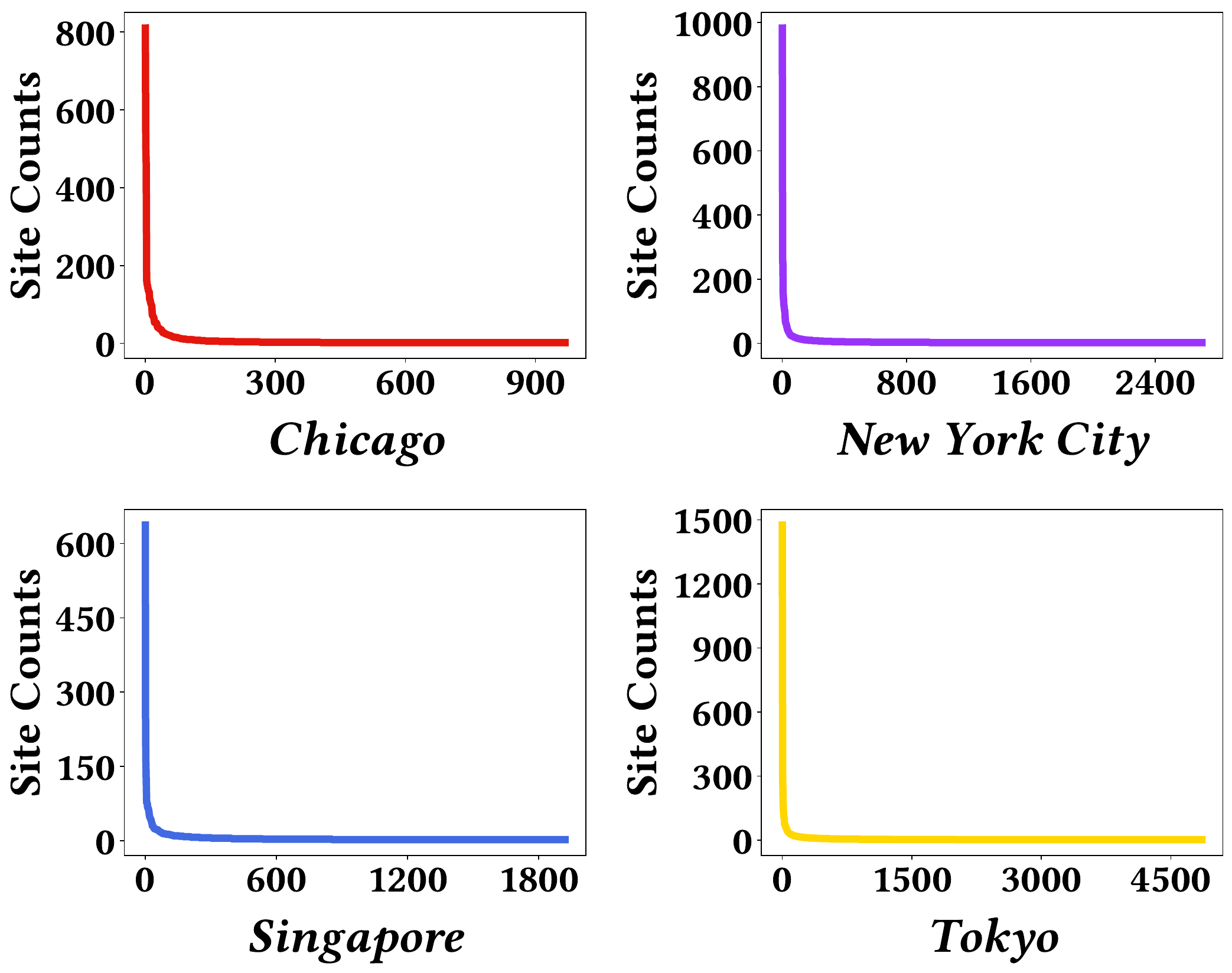}
    \caption{Distributions of site counts of brands in different cities. Every city shows the same trend that most of the POIs belong to an extremely small number of brands, which implies extremely severe long-tail problems on the brand side.}
    \label{fig:dist_brand}
\end{figure}

\begin{figure*}[h]
    \centering
    \subfigure[\textbf{Chicago}]{\includegraphics[width=0.28\linewidth]{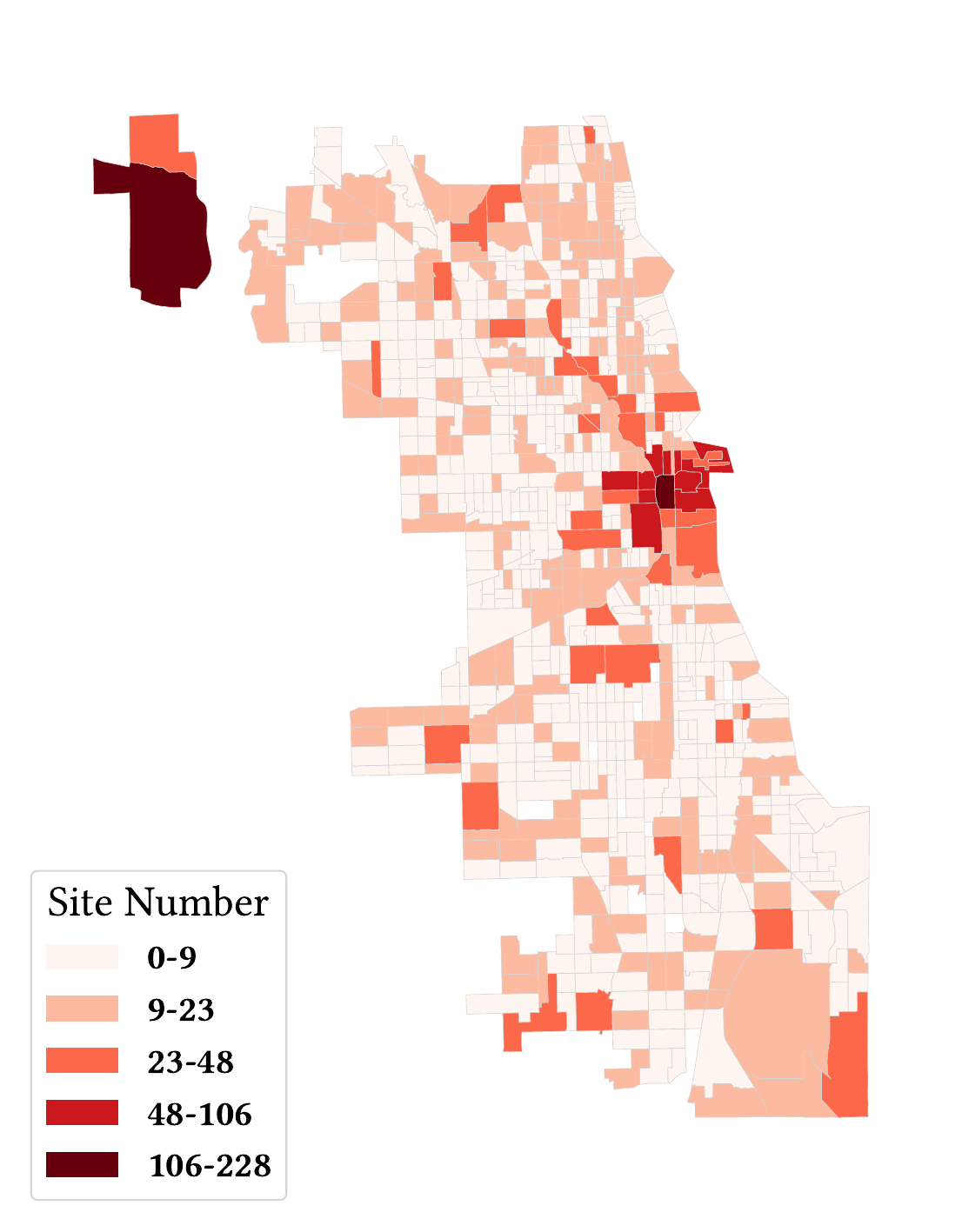}}
    \subfigure[\textbf{New York City}]{\includegraphics[width=0.37\linewidth]{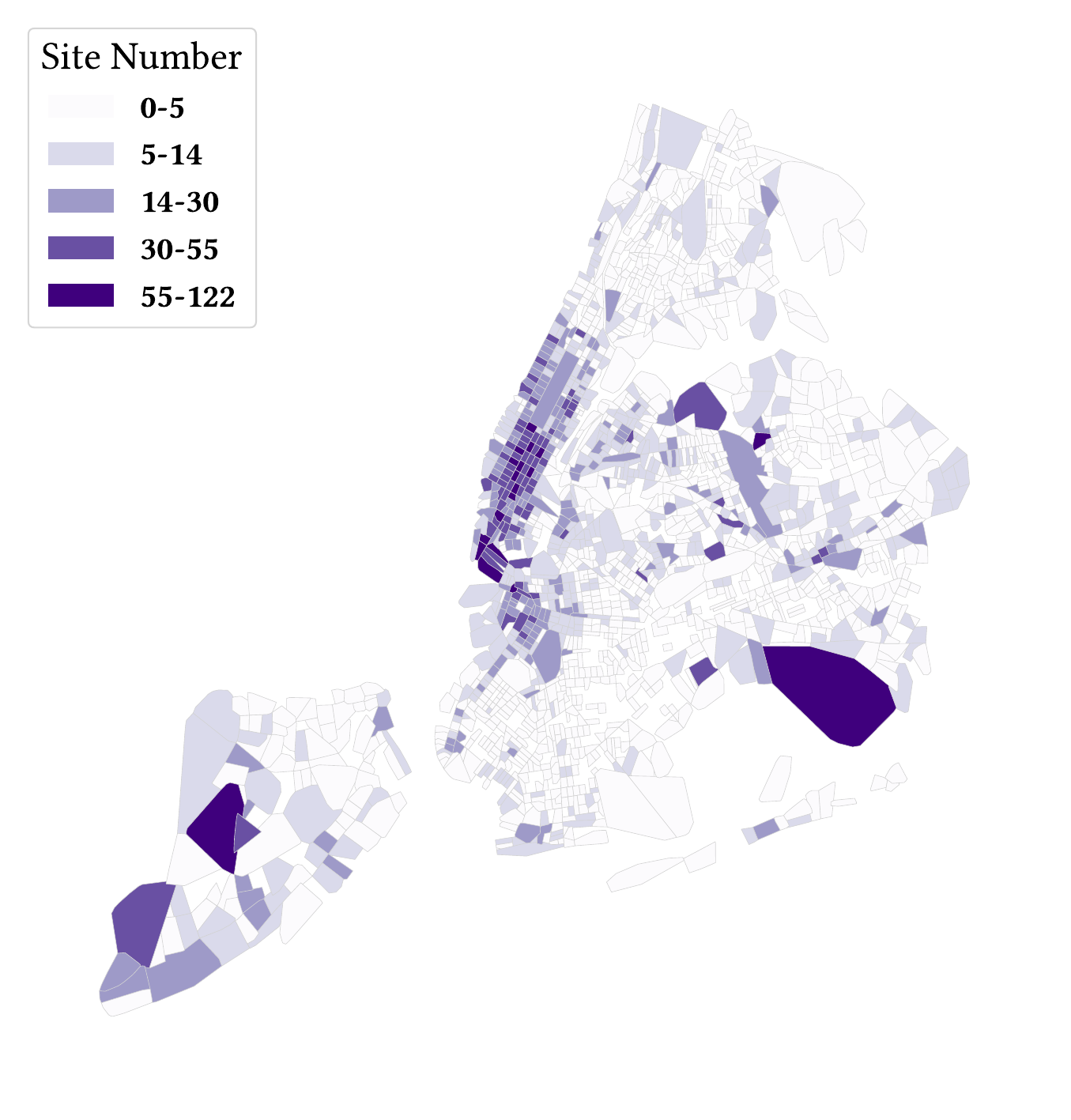}}
    \subfigure[\textbf{Singapore}]{\includegraphics[width=0.45\linewidth]{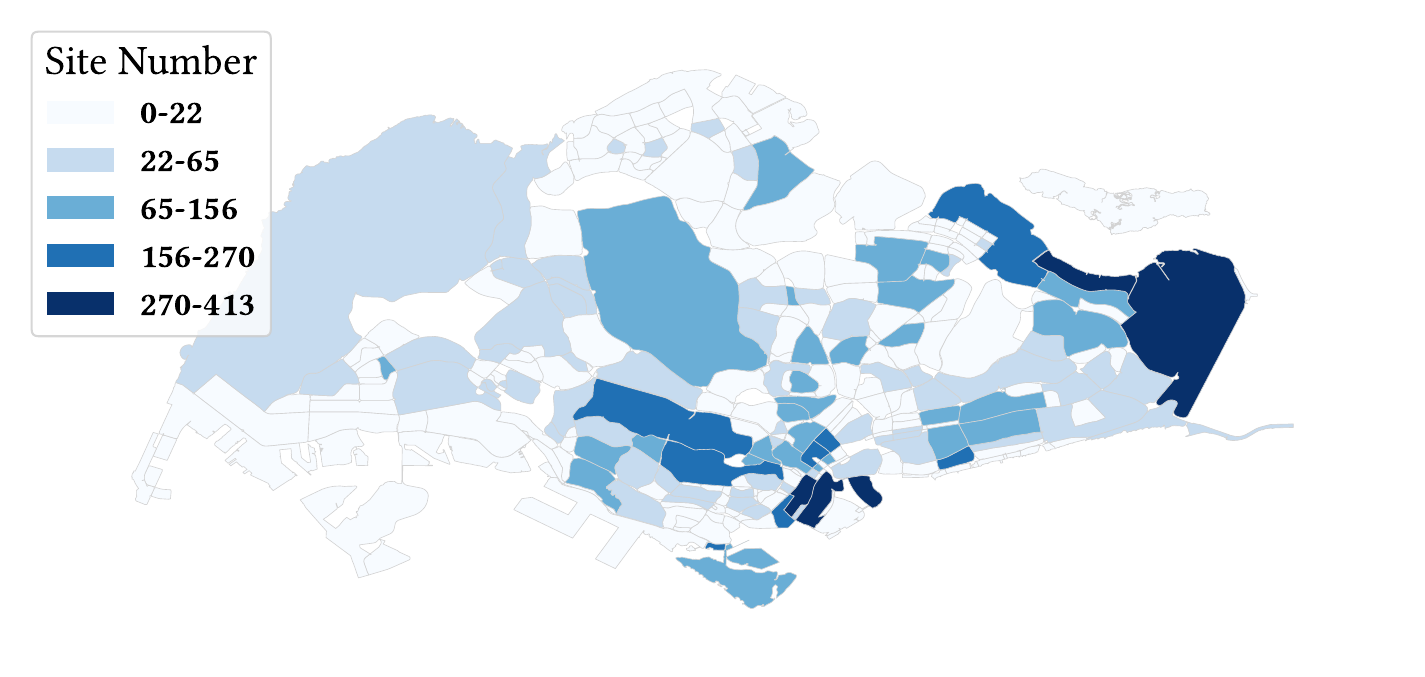}}
    \subfigure[\textbf{Tokyo}]{\includegraphics[width=0.32\linewidth]{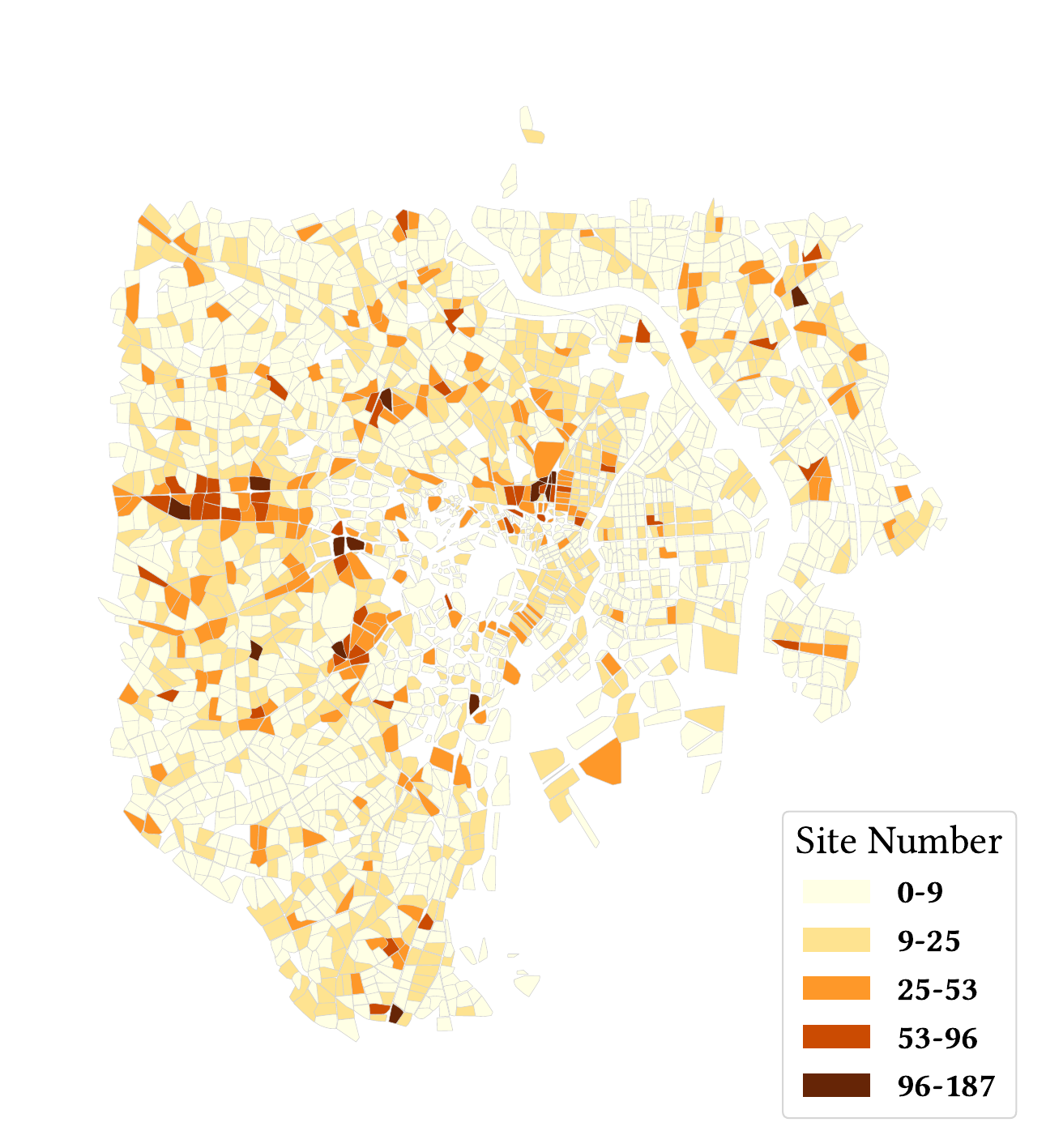}}
    \caption{Site distributions of regions in different cities. The uneven color distributions indicate that the site distributions are quite imbalanced on the region side.}
    \label{fig:dist_region}
\end{figure*}

Through the whole process of data collection and processing, we finally obtain the well-formulated and comprehensive \name dataset.
The detailed statistics are shown in Table~\ref{tab:dataset}.
In each city, we have thousands of brands and regions that serve the role of users and items in personalized recommendation systems, respectively.
Thus, the existing POIs that belong to specific brands and locate in specific regions serve the role of user-item interactions, whose amounts are about tens of thousand in each city.
Meanwhile, each POI has a series of hierarchical categories from about a hundred of types and locates in a specific business area (except for Tokyo).
Here we provide an example of a branch store of Starbucks in Tokyo as illustrated in Table~\ref{tab:example}.

In order to better facilitate the usage of \name, we further explore the characteristics of \name by analyzing the dataset distributions.
Due to the different urban planning and lifestyles of people, the distributions of site counts of categories are also different in each city.
From the comparison of distributions in Figure~\ref{fig:dist_category}, we can explore the highlights in these four metropolises.
For example, the people in Chicago and New York City are more likely to pay attention to the sports facilities, the choices of eating are more diverse in Singapore and the public transportation may be more convenient in Tokyo.

However, on the perspectives of brands and regions, the distributions are much more imbalanced.
As shown in Figure~\ref{fig:dist_brand}, the site counts of different brands are extremely imbalanced that the top 10\% of brands occupy over 50\% of sites.
This phenomenon indicates that the giant brands are actually dominating the commerce in every city.
Similarly, the uneven color distributions in Figure~\ref{fig:dist_region} show that the vast majority of POIs (up to 80\%) locate in several centres or sub-centres composed of a few regions while most regions have only a few POIs.
Such an inequality among different regions shows the different roles of the regions in urban planning, e.g., business areas are dense of POIs while residential areas are sparse.
All these distributions indicate the significant imbalance of \name, which is an important problem in utilizing the dataset for site recommendation.

On the basis of these statistics of \name, we conclude five unique characteristics of the site recommendation task compared with other top-N recommendation tasks:
\begin{itemize}[leftmargin=*]
    \item The data is quite \textbf{sparse}. Different from the user-oriented recommendation tasks, such as e-commerce or POI recommendation, the site recommendation task is brand-oriented. Due to the real-world concepts, the amounts of brands, regions and POIs are limited. Thus, each data point is very valuable and important.
    \item The \textbf{city-specific} characteristics are significant. It is necessary for site recommendation to effectively capture the latent essential of the city caused by urban planning and lifestyles. Meanwhile, the inter-city commonalities and differences will also play important roles in the researches of site recommendation, especially the researches on knowledge transfer across cities.
    \item The data distribution is extremely \textbf{imbalanced}. The significant imbalance may lead to undesirable predictions or other problems like popularity bias. Therefore, solving the imbalance problem is an important criterion in site recommendation.
    \item The \textbf{domain-specific features} are very crucial. Although there are rich domain-specific features in each domain, these features often have limited effects on the performance, such as the social~\cite{DBLP:conf/www/0002YZTZW19} and geographical~\cite{DBLP:conf/cikm/ChangJKK20} features in POI recommendation. However, due to the data rareness, the commercial and geographical features are crucial in site recommendation.
    \item The correlations are highly \textbf{complex}. While social relationships can be treated as positive correlations~\cite{DBLP:conf/www/Fan0LHZTY19} in social recommendation, the relationships between brands and between regions are much more complex. For example, two competitive brands may open stores in the same region and a brand may open stores in nearby regions, which is anti-intuitive but not unusual.
\end{itemize}

\section{Benchmark Experiments}
To better illustrate the significant importance of \name and deliver the facilitation for future research on site recommendation, we conduct a benchmark experiment.
In this section, we will report the experimental results with some widely-used baselines on \name under different scenarios as the benchmark.

\subsection{Experimental Settings}

\subsubsection{Dataset Split \& Evaluation Metrics}
Since the original \name is extremely imbalanced on both brands and regions because of the urban planning, we filter the dataset by 5-core setting on the brands (all the brands have possessed at least 5 POIs) in the benchmarking experiments.
Specifically, we randomly split the POIs of each brand with 70\%, 10\% and 20\% as training, validation and test sets, respectively.

To assess the model performance, we choose the widely-used standard evaluation metrics \textbf{Recall@20} and \textbf{nDCG@20} and regard all regions as candidates, i.e. all-ranking protocol.

\subsubsection{Baselines}
In order to explore the effectiveness of our proposed \name in practical application, we conduct experiments of site recommendation with several representative recommendation models as the benchmarks.
Since the existing site recommendation approaches have their own experimental settings and do not release their codes, they are not capable of directly being employed in our settings and we exclude them for comparison.
Specifically, the chosen models include machine learning models, collaborative filtering~\cite{DBLP:conf/www/SarwarKKR01} models, click-through rate (CTR) prediction~\cite{DBLP:conf/www/RichardsonDR07} models and graph-based models.

For machine learning models, we choose:
\begin{itemize}[leftmargin=*]
    \item \textbf{LR}~\cite{Hosmer1991AppliedLR}, namely logistic regression, is a simple yet effective model in classification.
    \item \textbf{GBDT}~\cite{Friedman2001GreedyFA}, namely gradient boosting decision tree, is an ensemble model with decision tree model as the backbone.
    \item \textbf{SVC}~\cite{DBLP:journals/coap/BredensteinerB99} employs support vector machine (SVM) to tackle the classification problem.
    \item \textbf{RankNet}~\cite{DBLP:conf/icml/BurgesSRLDHH05} is a famous learning to rank architecture in recommendation. Specifically, we utilize a two-layer neural network as the backbone here.
\end{itemize}

For collaborative filtering models, we choose:
\begin{itemize}[leftmargin=*]
    \item \textbf{MF-BPR}~\cite{DBLP:conf/uai/RendleFGS09} is a variant of Matrix Factorization (MF)~\cite{DBLP:journals/computer/KorenBV09} optimized by the Bayesian personalized ranking (BPR) loss.
    \item \textbf{NeuMF}~\cite{DBLP:conf/www/HeLZNHC17} combines neural network and Matrix Factorization (MF) for collaborative filtering with point-wise loss.
    \item \textbf{FISM}~\cite{DBLP:conf/kdd/KabburNK13} extends MF by aggregating the item embeddings of interacted items to represent the user via item similarity.
    \item \textbf{NAIS}~\cite{DBLP:journals/tkde/HeHSLJC18} introduces attention mechanism onto FISM to conduct weighted aggregation of items.
\end{itemize}

For CTR prediction models, we choose:
\begin{itemize}[leftmargin=*]
    \item \textbf{DNN}~\cite{DBLP:conf/recsys/CovingtonAS16} applies deep neural network to capture the complex interaction between features for CTR prediction.
    \item \textbf{Wide\&Deep}~\cite{DBLP:conf/recsys/Cheng0HSCAACCIA16} jointly utilizes linear transformation and DNN for CTR prediction.
    \item \textbf{DeepFM}~\cite{DBLP:conf/ijcai/GuoTYLH17} combines the factorization machine (FM) and DNN to model the first-, second- and high-order feature interactions.
    \item \textbf{xDeepFM}~\cite{DBLP:conf/kdd/LianZZCXS18} leverages the compressed interaction network (CIN) to achieve vector-wise feature interactions.
\end{itemize}

For graph-based models, we choose:
\begin{itemize}[leftmargin=*]
    \item \textbf{GC-MC}~\cite{DBLP:journals/corr/BergKW17} is a general graph neural network architecture for recommendation.
    \item \textbf{GraphRec}~\cite{DBLP:conf/www/Fan0LHZTY19} introduces graph neural network into social recommendation by aggregating the embeddings with social relationships. Here we remove the social aggregation component.
    \item \textbf{NGCF}~\cite{DBLP:conf/sigir/Wang0WFC19} namely Neural Graph Collaborative Filtering, conducts graph message passing on user-item interaction graph for recommendation.
    \item \textbf{LightGCN}~\cite{DBLP:conf/sigir/0001DWLZ020} simplifies the graph convolutional network with only neighborhood aggregation for collaborative filtering.
\end{itemize}

\begin{table*}[t]
    \caption{Benchmarking experimental results on OpenSiteRec. Best results in each category are highlighted in bold.}
    \centering
    \scalebox{1}
    {
        \begin{tabular}{l|cc|cc|cc|cc}
        \toprule
        \multirow{2}*{\textbf{Method}} & \multicolumn{2}{c}{\textbf{Chicago}} & \multicolumn{2}{c}{\textbf{New York City}} & \multicolumn{2}{c}{\textbf{Singapore}} & \multicolumn{2}{c}{\textbf{Tokyo}} \\
        & \textbf{Rec@20} & \textbf{nDCG@20} & \textbf{Rec@20} & \textbf{nDCG@20} & \textbf{Rec@20} & \textbf{nDCG@20} & \textbf{Rec@20} & \textbf{nDCG@20}\\
        \midrule
        LR & 0.1203 & 0.0868 & 0.0886 & 0.0655 & 0.1784 & 0.1336 & 0.0795 & 0.0594\\
        GBDT & 0.1203 & 0.0868 & 0.0886 & 0.0655 & 0.1784 & 0.1336 & 0.0795 & 0.0594\\
        SVC & 0.1203 & 0.0868 & 0.0886 & 0.0655 & 0.1784 & 0.1336 & 0.0795 & 0.0594\\
        RankNet & \textbf{0.2269} & \textbf{0.1427} & \textbf{0.1224} & \textbf{0.0654} & \textbf{0.4297} & \textbf{0.2271} & \textbf{0.1213} & \textbf{0.0667}\\
        \midrule
        MF-BPR & 0.2494 & 0.1465 & 0.1702 & 0.0917 & 0.4430 & 0.2351 & 0.1323 & \textbf{0.0781} \\
        NeuMF & 0.1942 & 0.1293 & 0.1200 & 0.0576 & 0.4289 & 0.2236 & 0.1225 & 0.0639\\
        FISM & \textbf{0.2547} & \textbf{0.1468} & \textbf{0.1745} & 0.0928 & \textbf{0.4583} & \textbf{0.2382} & \textbf{0.1343} & 0.0734\\
        NAIS & 0.2534 & 0.1432 & 0.1743 & \textbf{0.0964} & 0.4557 & 0.2380 & 0.1328 & 0.0774\\
        \midrule
        DNN & \textbf{0.1927} & \textbf{0.1311} & \textbf{0.1215} & \textbf{0.0568} & \textbf{0.4268} & \textbf{0.2204} & \textbf{0.1284} & \textbf{0.0648}\\
        Wide\&Deep & 0.1910 & 0.1284 & 0.1193 & 0.0526 & 0.4202 & 0.2174 & 0.1225 & 0.0647\\
        DeepFM & 0.1898 & 0.1254 & 0.1184 & 0.0531 & 0.4177 & 0.2182 & 0.1225 & \textbf{0.0648}\\
        xDeepFM & 0.1886 & 0.1225 & 0.1157 & 0.0515 & 0.4178 & 0.2180 & 0.1237 & 0.0647\\
        \midrule
        GC-MC & 0.2332 & 0.1657 & 0.1514 & 0.0513 & 0.4685 & 0.2317 & 0.1558 & 0.0884\\
        GraphRec & 0.2365 & 0.1640 & 0.1538 & 0.0550 & 0.4697 & 0.2293 & 0.1594 & 0.0905\\
        NGCF & 0.2866 & 0.1838 & 0.1920 & \textbf{0.1102} & 0.4929 & 0.2674 & 0.1619 & 0.1012\\
        LightGCN & \textbf{0.2875} & \textbf{0.1902} & \textbf{0.2087} & 0.1088 & \textbf{0.5013} & \textbf{0.2745} & \textbf{0.1751} & \textbf{0.1068}\\
        \bottomrule
        \end{tabular}
    }
    \label{tab:benchmark}
\end{table*}

\subsubsection{Implementation Details}
The experiments are implemented on the server with an Intel Xeon E5-2640 CPU, a 188GB RAM and two NVIDIA GeForce RTX 2080Ti GPUs.
According to the proposed problem definition, the brands and the regions actually serve the role of the users and the items, the categories and the business areas are treated as descriptive features and the site recommendation can be seen as a top-N recommendation task.
Specifically, for the traditional machine learning models, including LR, GDBT and SVC, we implement them with scikit-learn~\cite{scikit-learn} 1.0.2.
For other models that involve low-dimensional embeddings, we implement them with PyTorch~\cite{DBLP:conf/nips/PaszkeGMLBCKLGA19} 1.12.1 and set the embedding dimension as 100.
The model parameters are initialized with Xavier initialization and optimized by Adam~\cite{DBLP:journals/corr/KingmaB14}.
For all models, we tune hyper-parameters with the performance on the validation set via grid search.
The detailed implementation codes could be found at the git repository\footnote{https://github.com/HestiaSky/OpenSiteRec}.

\subsection{Benchmark Results}

The benchmark results of the baselines on \name are shown in Table~\ref{tab:benchmark}.
In order to deliver our insights for future research, we analyze the experimental results and summarize the following points:
\begin{itemize}[leftmargin=*]
    \item Traditional machine learning methods are not capable to handle the complex scenario of site recommendation. Although LR, GBDT and SVC are essentially different types of methods, they all converge to the same local optimal point and thus have the same performance. Given the situation with highly limited data but highly rich features, it is extremely difficult for the traditional machine learning methods not to over-fit to the training set. Thus, they fail to generalize to new POIs and are not suitable for site recommendation without additional mechanisms.
    \item The pair-wise loss is significantly better than the point-wise loss. As shown in the results, the models with pair-wise loss (i.e. BPR loss) including RankNet, MF-BPR, FISM, NAIS, NGCF, LightGCN, all outperform the models with point-wise loss (i.e. BCE loss) including NeuMF, DNN, Wide\&Deep, DeepFM, xDeepFM, GC-MC, GraphRec. This phenomenon indicates that the pair-wise loss is
    more suitable than the point-wise loss in site recommendation.
    \item The feature interaction has marginal effects on performance. From DNN to xDeepFM, the degree of feature interaction is increasing but the performance improvement is not significant. This may be because the correlations of features are either fully dependent (broad category and narrow category) or fully independent (brand category and geographical coordinate).
    \item The high-order interactions between brands and regions are crucial. Under the same condition of other factors, such as loss and feature interaction components, the graph-based models are substantially better than others. Since the data are highly sparse in site recommendation, high-order interactions are important to make correct predictions. Therefore, exploiting the graph representation learning techniques to better model the high-order interactions between brands and regions, especially the explicitly defined relations like \textit{Competitive}, are beneficial to obtain high performance in site recommendation.
\end{itemize}

\begin{figure*}
    \centering
    \includegraphics[width=0.9\linewidth]{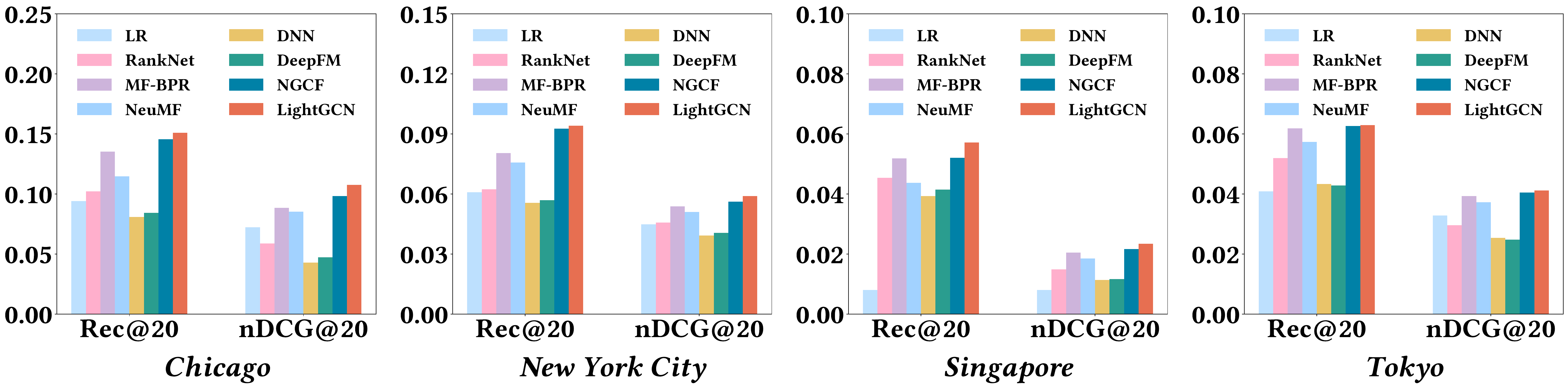}
    \caption{Experimental results of long-tail regions on OpenSiteRec.}
    \label{fig:lt_res}
\end{figure*}

\begin{table*}
    \caption{Comparison of the top brands of \textit{Fast food} (in red) and \textit{Cafe \& Dessert} (in blue) in different cities. The dark color denotes the more local special brand that hasn't entered all the other cities while the light color denotes the more international chain brand that has opened stores in every city.}
    \centering
    \scalebox{1}
    {
        \begin{tabular}{lr|lr|lr|lr}
            \toprule
            \multicolumn{2}{c|}{\textbf{Chicago}} & \multicolumn{2}{c|}{\textbf{New York City}} & \multicolumn{2}{c|}{\textbf{Singapore}} & \multicolumn{2}{c}{\textbf{Tokyo}} \\
            \textbf{Brand} & \textbf{Portion} & \textbf{Brand} & \textbf{Portion} & \textbf{Brand} & \textbf{Portion} & \textbf{Brand} & \textbf{Portion} \\
            \midrule
            \textcolor[rgb]{0.3,0.6,1}{Dunkin’}	& 6.15\% & \textcolor[rgb]{0.3,0.6,1}{Dunkin’} & 4.24\% & \textcolor[rgb]{0.3,0.6,1}{Starbucks} & 3.33\% & \textcolor[rgb]{0,0,1}{\textbf{Doutor}} & 2.65\%  \\
            \textcolor[rgb]{1,0.5,0.5}{Subway} & 5.32\% & \textcolor[rgb]{0.3,0.6,1}{Starbucks} & 3.17\% & \textcolor[rgb]{1,0.5,0.5}{McDonald's} & 3.01\% & \textcolor[rgb]{0.3,0.6,1}{Starbucks} & 1.98\% \\
            \textcolor[rgb]{0.3,0.6,1}{Starbucks} & 5.17\% & \textcolor[rgb]{1,0.5,0.5}{McDonald's} & 1.98\% & \textcolor[rgb]{1,0,0}{\textbf{Toast Box}} & 1.98\% & \textcolor[rgb]{1,0.5,0.5}{McDonald's} & 1.80\%\\
            \textcolor[rgb]{1,0.5,0.5}{McDonald’s}	& 4.26\% &	\textcolor[rgb]{1,0.5,0.5}{Subway} & 1.63\% & \textcolor[rgb]{1,0.5,0.5}{KFC} & 1.79\% & \textcolor[rgb]{1,0.1,0.1}{\textbf{Matsuya}} & 1.77\% \\
            \textcolor[rgb]{1,0.5,0.5}{Burger King} & 1.85\% & \textcolor[rgb]{1,0.5,0.5}{Burger King} & 0.98\% & \textcolor[rgb]{1,0.5,0.5}{Subway} & 1.77\% & \textcolor[rgb]{1,0,0}{\textbf{Sukiya}} & 1.18\% \\
            \textcolor[rgb]{0,0,1}{\textbf{Baskin-Robbins}} & 1.55\% & \textcolor[rgb]{1,0.5,0.5}{Popeyes} & 0.83\% & \textcolor[rgb]{0,0,1}{\textbf{Coffee Bean}} & 1.77\% & \textcolor[rgb]{1,0.1,0.1}{\textbf{MOS Burger}} & 1.17\%\\
            \textcolor[rgb]{1,0.5,0.5}{Popeyes} & 1.36\% & \textcolor[rgb]{1,0,0}{\textbf{Chipotle}} & 0.83\% & \textcolor[rgb]{1,0,0}{\textbf{Ya Kun Kaya Toast}} & 1.54\% & \textcolor[rgb]{0,0,1}{\textbf{Tully's Coffee}} & 1.17\% \\
            \textcolor[rgb]{1,0,0}{\textbf{Potbelly}} & 1.36\% & \textcolor[rgb]{1,0.5,0.5}{Domino's} & 0.54\% & \textcolor[rgb]{0,0,1}{\textbf{Kopitiam}} & 1.10\% & \textcolor[rgb]{1,0,0}{\textbf{Yoshinoya}} & 1.02\% \\
            \bottomrule
        \end{tabular}
    }
    \label{tab:case}
\end{table*}

\begin{figure*}
    \centering
    \subfigure[\textbf{Chicago}]{\includegraphics[width=0.28\linewidth]{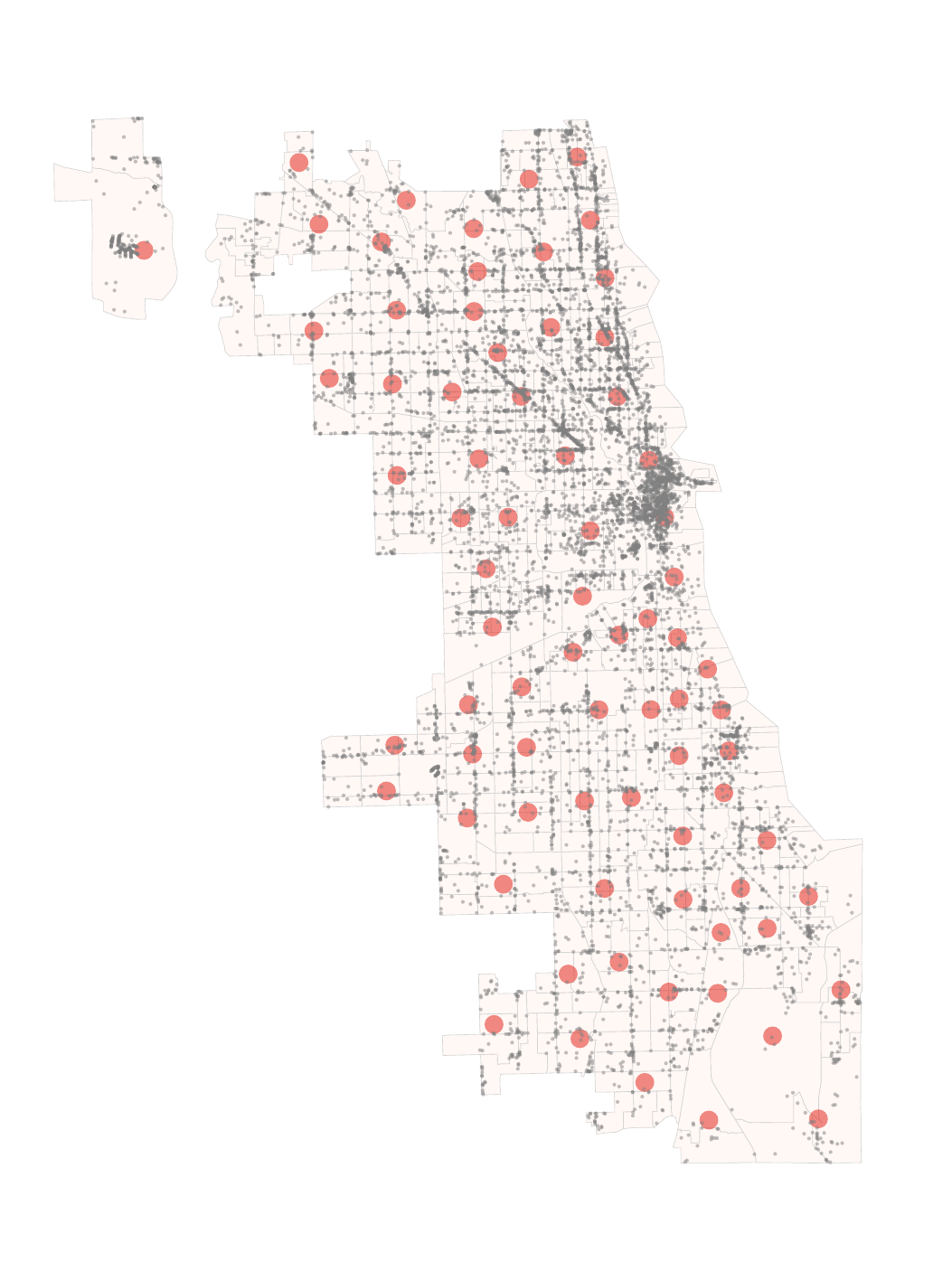}}
    \subfigure[\textbf{New York City}]{\includegraphics[width=0.37\linewidth]{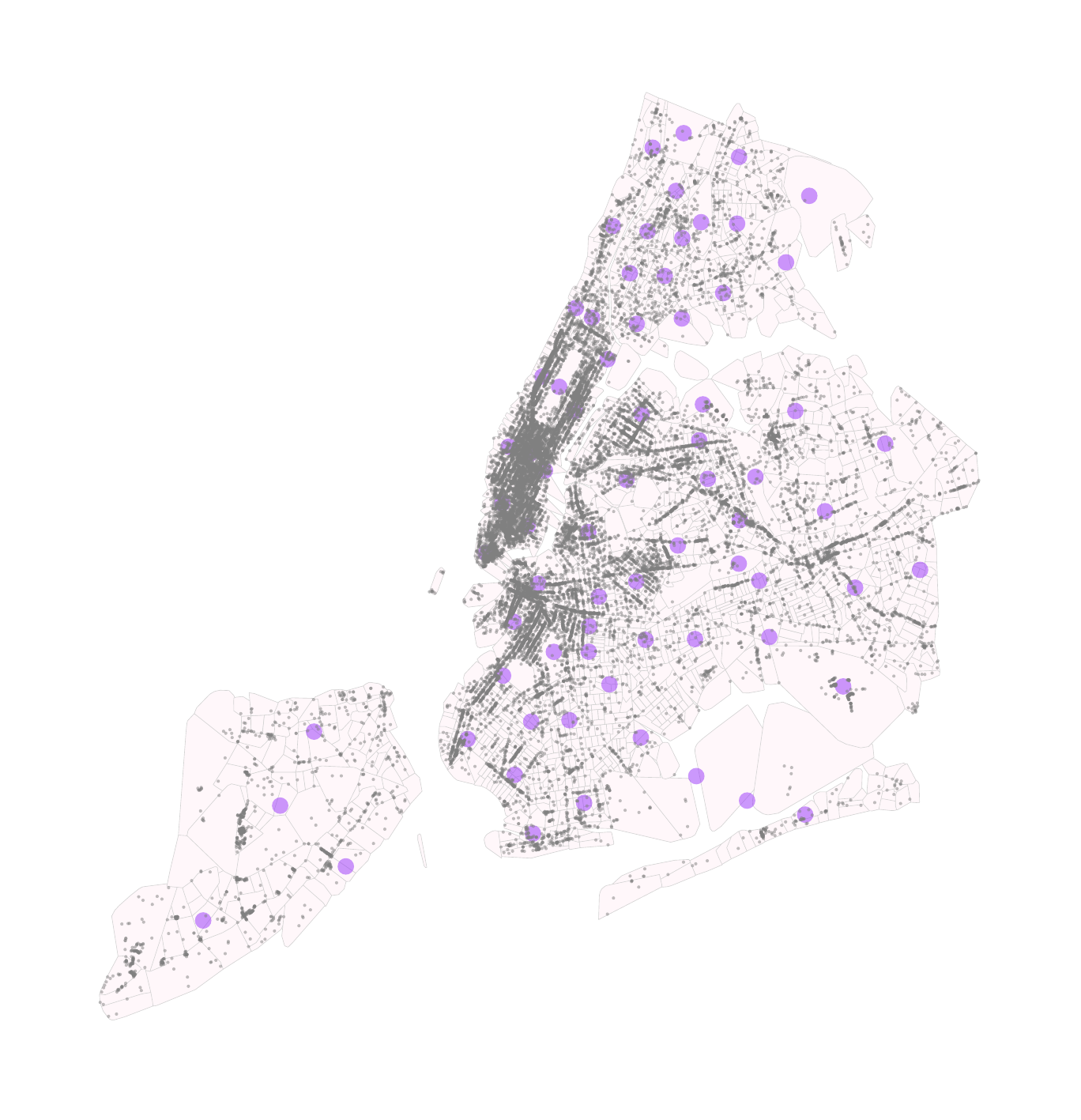}}
    \subfigure[\textbf{Singapore}]{\includegraphics[width=0.45\linewidth]{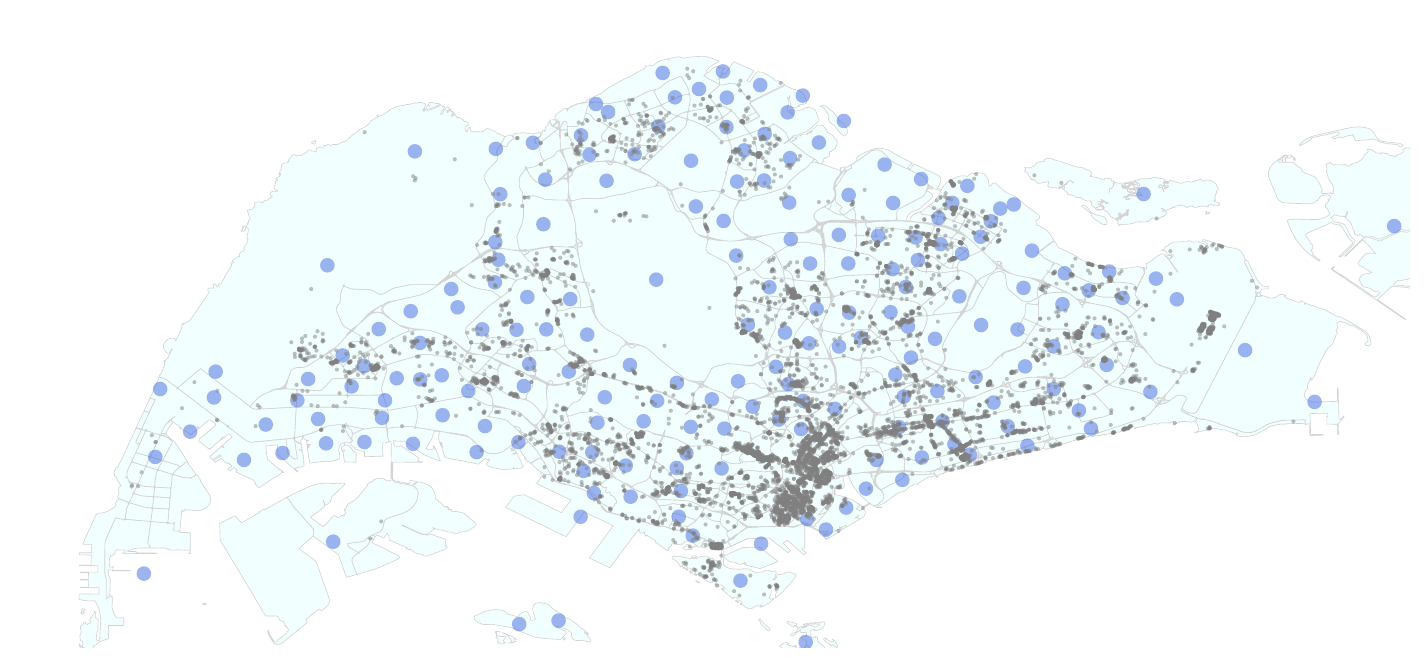}}
    \subfigure[\textbf{Tokyo}]{\includegraphics[width=0.32\linewidth]{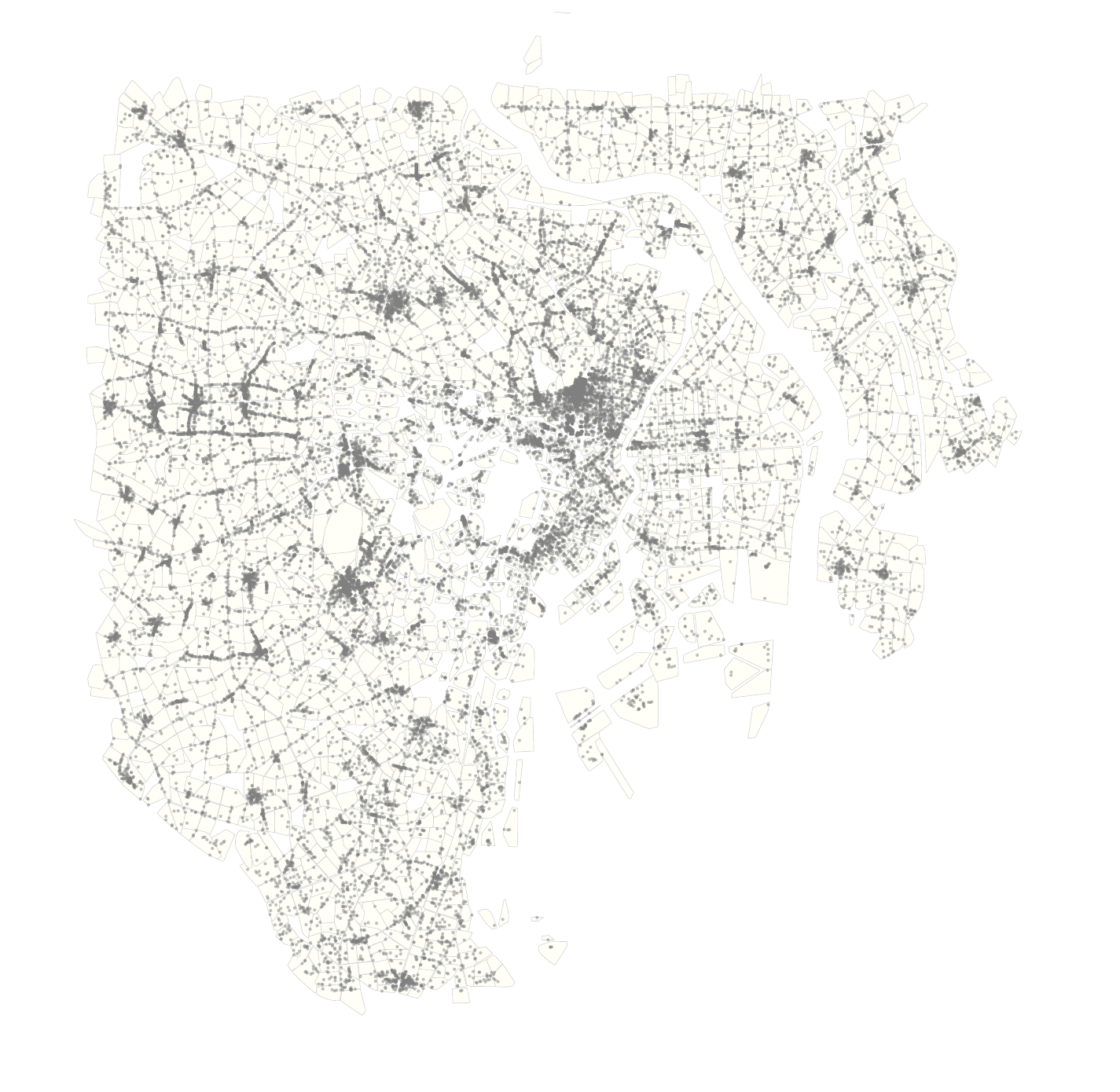}}
    \caption{Visualization of the planned business area of different cities with POI distributions. Specifically, the government of Tokyo has not released the official business area planning so there is no business area in the subfigure.}
    \label{fig:dist}
\end{figure*}

\subsection{Long-tail Scenario}

Since the imbalance problem is severe in site recommendation, we conduct an additional experiment to evaluate some representative baseline models, including LR, RankNet, MF-BPR, NeuMF, DNN, DeepFM, NGCF and LightGCN, under the long-tail scenario.
Specifically, we regard the bottom 90\% of regions with fewer POIs as the long-tail regions which totally occupy less than 50\% of POIs and are only for testing.

As shown in Figure~\ref{fig:lt_res}, the performances of the baseline models drop dramatically on these long-tail regions, which implies a big challenge in site recommendation.
Meanwhile, we can also find that the advantage of graph-based models is not as significant as it under the vanilla scenario, which is an important point in designing the model architecture.
From the perspective of the cities, the greater drop in performance from the vanilla scenario to the long-tail scenario indicates a higher degree of centrality in urban planning.
According to the results, \textit{Chicago} and \textit{Singapore} have a higher degree of centrality than \textit{New York City} and \textit{Tokyo}, which means the giant brands have more dominant positions in \textit{Chicago} and \textit{Singapore}.
However, to better achieve fairness, which is especially crucial to provide forward-looking recommendation results to promote urban development, addressing this extreme data imbalance issue is also an important and valuable research topic in site recommendation.

\section{Potential Applications}
Besides site recommendation, our \name also support many other potential applications, including spatial object recommendation~\cite{DBLP:conf/sigir/LuoZBLCYLX20}, transportation demand prediction~\cite{DBLP:conf/aaai/YeSDF021,DBLP:conf/aaai/LiuWZLD021}, electric vehicle charging recommendation~\cite{DBLP:conf/www/ZhangLWXXDX21,DBLP:conf/kdd/00030HG022} and high-potential startup detection~\cite{DBLP:conf/kdd/ZhangZYX21}, which demonstrates the strong applicability of \name.
In this section, we discuss the significance and show the feasibility via case study and visualization of two potential tasks: brand entry forecasting~\cite{Seenivasan2016CompetitiveEO,Maesen2022TheIO} and business area planning~\cite{Yu2015TheAA}.

\subsection{Brand Entry Forecasting}

People in different cities have different preferences for brands due to culture, history, lifestyle and other reasons~\cite{Lehmann1994ContextEN,Pan1993TheIO}.
These factors along with the commercial strategies of brands result in the different brand distributions in different cities~\cite{Pauwels2004WhoBF}.
Typically, some brands are more international that have been spread all over the world while some brands are more local that only open stores at a few cities.
Obviously, all these international brands started from local brands with decades of expansion to reach current standings.
Therefore, many brands that are very popular in local now have a great potential to expand to other cities around the world~\cite{Carpenter1990CompetitiveSF}.
A feasible way is to exploit the existing brands in different cities to mine the brands with a high probability to succeed in other cities~\cite{Zhang2017DemandFA,DBLP:conf/kdd/ZhangZYX21}, i.e. brand entry forecasting.
Since our proposed \name provides plenty of brands in four world-class metropolises, it is credible to carry out such research with \name.

In order to demonstrate the applicability of our \name to support the brand entry forecasting, we also present a case study by ranking the brands of \textit{Fast Food} and \textit{Cafe \& Dessert} in different cities.
As illustrated in Table~\ref{tab:case}, many popular international brands like \textit{Starbucks} have multiple similar brands that are popular in local, such as \textit{Coffee Bean} in \textit{Singapore} and \textit{Doutor} in \textit{Tokyo}.
The fact that these local brands have the ability to pose a position in the fierce commercial competition with the international giant brands not only demonstrates their success in commerce so far, but also indicates their adequate competitiveness to success in other places in the future.
Therefore, it is fair to believe that these brands have a high potential to enter other cities and our \name is capable to provide valuable information for this research topic.

\subsection{Business Area Planning}

To form a strong scale effect, planning business area~\cite{Yu2015TheAA} (i.e. central business district) is crucial in the development of the city.
Generally, business area planning takes many factors into account, such as population, traffic convenience and surrounding environments.
Since the POI distributions and the geographical information already indicate these factors implicitly~\cite{DBLP:conf/kdd/KaramshukNSNM13}, we deem that \name is quite beneficial to support business area planning.

To verify this idea, we visualize the planned business area (except for Tokyo, whose government has not released the official business area planning) and POI distributions for comparison.
From Figure~\ref{fig:dist}, we can see that the distributions of business areas are relatively uniform in the cities and have high coincidence with the traffic hubs.
Such a regular pattern is consistent with the POI distributions and it is also true that the more concentrated the POIs are, the more business districts are planned.
Our \name provides tens of thousands of POIs along with their categories and other characteristics.
By analyzing these characteristics of all the POIs in an area, it is to effectively grasp the essential factors of the area.
For example, the areas with high density of shopping sites or the areas serving as transportation hubs will be more possible and reasonable to be planned as business areas.
Therefore, applying \name to conduct research in business area planning is feasible and reliable.

\section{Discussion \& Conclusion}
In this paper, we collect, construct and release the first open dataset \name for site recommendation, which consists of multi-source data from four international metropolises and is comprehensive to support the following research.
Specifically, \name leverages a heterogeneous graph schema with the entities to represent the different types of real-world concepts, including category, brand, POI, business area and region, and the relations to denote the corresponding commercial and geographical relationships.

Overall, site recommendation is an important and beneficial information retrieval task for the development of brands or institutions in real-world applications, which has been established for decades.
However, it was until recent years that the research on site recommendation has made much progress, but it is still very slow compared with other rapidly developing areas, such as POI recommendation.
Despite the success of recent approaches in site recommendation, none of them have released their datasets, which brings inconvenience for the researchers and is harmful to the research in this area.
Meanwhile, most of the existing approaches only focus on a small scope of site recommendation, yielding limited significance and impact.
Our \name provides comprehensive and variety of information across multiple cities to better support more practical and valuable research.

To verify the applicability of our \name and the suitability of the existing recommendation models on site recommendation task, we conduct benchmarking experiments of totally 16 representative baseline models on \name.
The experimental results fully demonstrate the great support for site recommendation and the detailed analysis also provides some insights in developing more advanced models.
Furthermore, we explore the potential research directions and deliver toy experiments of urban computing, including brand entry forecasting and business area planning, which indicates the applicability of our \name in supporting various real-world applications.

Unfortunately, there are still some potential limitations of \name for now.
The greatest limitation consists in the lack of temporal dimension, which means \name collects the data at a specific time without variation from urban development.
Therefore, our \name contains the real-world data but not necessarily the best options of regions for the brands.
A better solution considering the temporal dimension is to collect data and construct dataset at multiple time points on annually, quarterly, or even monthly.
However, due to the requirements of original data and the high workload, we can not afford to achieve it at present and it is left for future work.
In addition, \name currently contains four metropolises, which are still limited.
We have planned to expand it by adding more cities to better support the research of site recommendation.

In conclusion, site recommendation is is still an underestimated topic considering its significance in modern business and its predicted impacts.
Therefore, we believe that the emergence of our \name dataset will strongly promote the research on it and help the development of business intelligence in the next few years.


\bibliographystyle{ACM-Reference-Format}
\bibliography{custom}


\begin{thebibliography}{71}


\ifx \showCODEN    \undefined \def \showCODEN     #1{\unskip}     \fi
\ifx \showDOI      \undefined \def \showDOI       #1{#1}\fi
\ifx \showISBNx    \undefined \def \showISBNx     #1{\unskip}     \fi
\ifx \showISBNxiii \undefined \def \showISBNxiii  #1{\unskip}     \fi
\ifx \showISSN     \undefined \def \showISSN      #1{\unskip}     \fi
\ifx \showLCCN     \undefined \def \showLCCN      #1{\unskip}     \fi
\ifx \shownote     \undefined \def \shownote      #1{#1}          \fi
\ifx \showarticletitle \undefined \def \showarticletitle #1{#1}   \fi
\ifx \showURL      \undefined \def \showURL       {\relax}        \fi
\providecommand\bibfield[2]{#2}
\providecommand\bibinfo[2]{#2}
\providecommand\natexlab[1]{#1}
\providecommand\showeprint[2][]{arXiv:#2}

\bibitem[\protect\citeauthoryear{Berman and Krass}{Berman and Krass}{2002}]%
        {DBLP:journals/cor/BermanK02}
\bibfield{author}{\bibinfo{person}{Oded Berman} {and} \bibinfo{person}{Dmitry
  Krass}.} \bibinfo{year}{2002}\natexlab{}.
\newblock \showarticletitle{The generalized maximal covering location problem}.
\newblock \bibinfo{journal}{\emph{Comput. Oper. Res.}} \bibinfo{volume}{29},
  \bibinfo{number}{6} (\bibinfo{year}{2002}), \bibinfo{pages}{563--581}.
\newblock


\bibitem[\protect\citeauthoryear{Bredensteiner and Bennett}{Bredensteiner and
  Bennett}{1999}]%
        {DBLP:journals/coap/BredensteinerB99}
\bibfield{author}{\bibinfo{person}{Erin~J. Bredensteiner} {and}
  \bibinfo{person}{Kristin~P. Bennett}.} \bibinfo{year}{1999}\natexlab{}.
\newblock \showarticletitle{Multicategory Classification by Support Vector
  Machines}.
\newblock \bibinfo{journal}{\emph{Comput. Optim. Appl.}} \bibinfo{volume}{12},
  \bibinfo{number}{1-3} (\bibinfo{year}{1999}), \bibinfo{pages}{53--79}.
\newblock


\bibitem[\protect\citeauthoryear{Burges, Shaked, Renshaw, Lazier, Deeds,
  Hamilton, and Hullender}{Burges et~al\mbox{.}}{2005}]%
        {DBLP:conf/icml/BurgesSRLDHH05}
\bibfield{author}{\bibinfo{person}{Christopher J.~C. Burges},
  \bibinfo{person}{Tal Shaked}, \bibinfo{person}{Erin Renshaw},
  \bibinfo{person}{Ari Lazier}, \bibinfo{person}{Matt Deeds},
  \bibinfo{person}{Nicole Hamilton}, {and} \bibinfo{person}{Gregory~N.
  Hullender}.} \bibinfo{year}{2005}\natexlab{}.
\newblock \showarticletitle{Learning to rank using gradient descent}. In
  \bibinfo{booktitle}{\emph{Machine Learning, Proceedings of the Twenty-Second
  International Conference {(ICML} 2005), Bonn, Germany, August 7-11, 2005}}
  \emph{(\bibinfo{series}{{ACM} International Conference Proceeding Series})},
  Vol.~\bibinfo{volume}{119}. \bibinfo{publisher}{{ACM}},
  \bibinfo{pages}{89--96}.
\newblock


\bibitem[\protect\citeauthoryear{Carpenter and Nakamoto}{Carpenter and
  Nakamoto}{1990}]%
        {Carpenter1990CompetitiveSF}
\bibfield{author}{\bibinfo{person}{Gregory~S. Carpenter} {and}
  \bibinfo{person}{Kent Nakamoto}.} \bibinfo{year}{1990}\natexlab{}.
\newblock \showarticletitle{Competitive Strategies for Late Entry into a Market
  with a Dominant Brand}.
\newblock \bibinfo{journal}{\emph{Management Science}}  \bibinfo{volume}{36}
  (\bibinfo{year}{1990}), \bibinfo{pages}{1268--1278}.
\newblock


\bibitem[\protect\citeauthoryear{Chang, Jang, Kim, and Kang}{Chang
  et~al\mbox{.}}{2020}]%
        {DBLP:conf/cikm/ChangJKK20}
\bibfield{author}{\bibinfo{person}{Buru Chang}, \bibinfo{person}{Gwanghoon
  Jang}, \bibinfo{person}{Seoyoon Kim}, {and} \bibinfo{person}{Jaewoo Kang}.}
  \bibinfo{year}{2020}\natexlab{}.
\newblock \showarticletitle{Learning Graph-Based Geographical Latent
  Representation for Point-of-Interest Recommendation}. In
  \bibinfo{booktitle}{\emph{{CIKM} '20: The 29th {ACM} International Conference
  on Information and Knowledge Management, Virtual Event, Ireland, October
  19-23, 2020}}. \bibinfo{publisher}{{ACM}}, \bibinfo{pages}{135--144}.
\newblock


\bibitem[\protect\citeauthoryear{Chen, Zhang, Pan, Ma, Yang, Kushlev, Zhang,
  and Li}{Chen et~al\mbox{.}}{2015}]%
        {DBLP:conf/huc/ChenZPMYKZL15}
\bibfield{author}{\bibinfo{person}{Longbiao Chen}, \bibinfo{person}{Daqing
  Zhang}, \bibinfo{person}{Gang Pan}, \bibinfo{person}{Xiaojuan Ma},
  \bibinfo{person}{Dingqi Yang}, \bibinfo{person}{Kostadin Kushlev},
  \bibinfo{person}{Wangsheng Zhang}, {and} \bibinfo{person}{Shijian Li}.}
  \bibinfo{year}{2015}\natexlab{}.
\newblock \showarticletitle{Bike sharing station placement leveraging
  heterogeneous urban open data}. In \bibinfo{booktitle}{\emph{Proceedings of
  the 2015 {ACM} International Joint Conference on Pervasive and Ubiquitous
  Computing, UbiComp 2015, Osaka, Japan, September 7-11, 2015}}.
  \bibinfo{publisher}{{ACM}}, \bibinfo{pages}{571--575}.
\newblock


\bibitem[\protect\citeauthoryear{Cheng, Koc, Harmsen, Shaked, Chandra, Aradhye,
  Anderson, Corrado, Chai, Ispir, Anil, Haque, Hong, Jain, Liu, and Shah}{Cheng
  et~al\mbox{.}}{2016}]%
        {DBLP:conf/recsys/Cheng0HSCAACCIA16}
\bibfield{author}{\bibinfo{person}{Heng{-}Tze Cheng}, \bibinfo{person}{Levent
  Koc}, \bibinfo{person}{Jeremiah Harmsen}, \bibinfo{person}{Tal Shaked},
  \bibinfo{person}{Tushar Chandra}, \bibinfo{person}{Hrishi Aradhye},
  \bibinfo{person}{Glen Anderson}, \bibinfo{person}{Greg Corrado},
  \bibinfo{person}{Wei Chai}, \bibinfo{person}{Mustafa Ispir},
  \bibinfo{person}{Rohan Anil}, \bibinfo{person}{Zakaria Haque},
  \bibinfo{person}{Lichan Hong}, \bibinfo{person}{Vihan Jain},
  \bibinfo{person}{Xiaobing Liu}, {and} \bibinfo{person}{Hemal Shah}.}
  \bibinfo{year}{2016}\natexlab{}.
\newblock \showarticletitle{Wide {\&} Deep Learning for Recommender Systems}.
  In \bibinfo{booktitle}{\emph{Proceedings of the 1st Workshop on Deep Learning
  for Recommender Systems, DLRS@RecSys 2016, Boston, MA, USA, September 15,
  2016}}. \bibinfo{publisher}{{ACM}}, \bibinfo{pages}{7--10}.
\newblock


\bibitem[\protect\citeauthoryear{Church and Murray}{Church and Murray}{2008}]%
        {Church2008BusinessSS}
\bibfield{author}{\bibinfo{person}{Richard~L. Church} {and}
  \bibinfo{person}{Alan~T. Murray}.} \bibinfo{year}{2008}\natexlab{}.
\newblock \showarticletitle{Business Site Selection, Location Analysis and
  GIS}.
\newblock


\bibitem[\protect\citeauthoryear{Cohen, Ravikumar, and Fienberg}{Cohen
  et~al\mbox{.}}{2003}]%
        {DBLP:conf/ijcai/CohenRF03}
\bibfield{author}{\bibinfo{person}{William~W. Cohen}, \bibinfo{person}{Pradeep
  Ravikumar}, {and} \bibinfo{person}{Stephen~E. Fienberg}.}
  \bibinfo{year}{2003}\natexlab{}.
\newblock \showarticletitle{A Comparison of String Distance Metrics for
  Name-Matching Tasks}. In \bibinfo{booktitle}{\emph{Proceedings of {IJCAI-03}
  Workshop on Information Integration on the Web (IIWeb-03), August 9-10, 2003,
  Acapulco, Mexico}}. \bibinfo{pages}{73--78}.
\newblock


\bibitem[\protect\citeauthoryear{Covington, Adams, and Sargin}{Covington
  et~al\mbox{.}}{2016}]%
        {DBLP:conf/recsys/CovingtonAS16}
\bibfield{author}{\bibinfo{person}{Paul Covington}, \bibinfo{person}{Jay
  Adams}, {and} \bibinfo{person}{Emre Sargin}.}
  \bibinfo{year}{2016}\natexlab{}.
\newblock \showarticletitle{Deep Neural Networks for YouTube Recommendations}.
  In \bibinfo{booktitle}{\emph{Proceedings of the 10th {ACM} Conference on
  Recommender Systems, Boston, MA, USA, September 15-19, 2016}}.
  \bibinfo{publisher}{{ACM}}, \bibinfo{pages}{191--198}.
\newblock


\bibitem[\protect\citeauthoryear{Fan, Ma, Li, He, Zhao, Tang, and Yin}{Fan
  et~al\mbox{.}}{2019}]%
        {DBLP:conf/www/Fan0LHZTY19}
\bibfield{author}{\bibinfo{person}{Wenqi Fan}, \bibinfo{person}{Yao Ma},
  \bibinfo{person}{Qing Li}, \bibinfo{person}{Yuan He},
  \bibinfo{person}{Yihong~Eric Zhao}, \bibinfo{person}{Jiliang Tang}, {and}
  \bibinfo{person}{Dawei Yin}.} \bibinfo{year}{2019}\natexlab{}.
\newblock \showarticletitle{Graph Neural Networks for Social Recommendation}.
  In \bibinfo{booktitle}{\emph{The World Wide Web Conference, {WWW} 2019, San
  Francisco, CA, USA, May 13-17, 2019}}. \bibinfo{publisher}{{ACM}},
  \bibinfo{pages}{417--426}.
\newblock


\bibitem[\protect\citeauthoryear{Friedman}{Friedman}{2001}]%
        {Friedman2001GreedyFA}
\bibfield{author}{\bibinfo{person}{Jerome~H. Friedman}.}
  \bibinfo{year}{2001}\natexlab{}.
\newblock \showarticletitle{Greedy function approximation: A gradient boosting
  machine.}
\newblock \bibinfo{journal}{\emph{Annals of Statistics}}  \bibinfo{volume}{29}
  (\bibinfo{year}{2001}), \bibinfo{pages}{1189--1232}.
\newblock


\bibitem[\protect\citeauthoryear{Guo, Li, Zheng, Wang, and Yu}{Guo
  et~al\mbox{.}}{2017a}]%
        {DBLP:journals/imwut/GuoLZWY17}
\bibfield{author}{\bibinfo{person}{Bin Guo}, \bibinfo{person}{Jing Li},
  \bibinfo{person}{Vincent~W. Zheng}, \bibinfo{person}{Zhu Wang}, {and}
  \bibinfo{person}{Zhiwen Yu}.} \bibinfo{year}{2017}\natexlab{a}.
\newblock \showarticletitle{CityTransfer: Transferring Inter- and Intra-City
  Knowledge for Chain Store Site Recommendation based on Multi-Source Urban
  Data}.
\newblock \bibinfo{journal}{\emph{Proc. {ACM} Interact. Mob. Wearable
  Ubiquitous Technol.}} \bibinfo{volume}{1}, \bibinfo{number}{4}
  (\bibinfo{year}{2017}), \bibinfo{pages}{135:1--135:23}.
\newblock


\bibitem[\protect\citeauthoryear{Guo, Tang, Ye, Li, and He}{Guo
  et~al\mbox{.}}{2017b}]%
        {DBLP:conf/ijcai/GuoTYLH17}
\bibfield{author}{\bibinfo{person}{Huifeng Guo}, \bibinfo{person}{Ruiming
  Tang}, \bibinfo{person}{Yunming Ye}, \bibinfo{person}{Zhenguo Li}, {and}
  \bibinfo{person}{Xiuqiang He}.} \bibinfo{year}{2017}\natexlab{b}.
\newblock \showarticletitle{DeepFM: {A} Factorization-Machine based Neural
  Network for {CTR} Prediction}. In \bibinfo{booktitle}{\emph{Proceedings of
  the Twenty-Sixth International Joint Conference on Artificial Intelligence,
  {IJCAI} 2017, Melbourne, Australia, August 19-25, 2017}}.
  \bibinfo{publisher}{ijcai.org}, \bibinfo{pages}{1725--1731}.
\newblock


\bibitem[\protect\citeauthoryear{Haklay and Weber}{Haklay and Weber}{2008}]%
        {DBLP:journals/pervasive/HaklayW08}
\bibfield{author}{\bibinfo{person}{Mordechai~(Muki) Haklay} {and}
  \bibinfo{person}{Patrick Weber}.} \bibinfo{year}{2008}\natexlab{}.
\newblock \showarticletitle{OpenStreetMap: User-Generated Street Maps}.
\newblock \bibinfo{journal}{\emph{{IEEE} Pervasive Comput.}}
  \bibinfo{volume}{7}, \bibinfo{number}{4} (\bibinfo{year}{2008}),
  \bibinfo{pages}{12--18}.
\newblock


\bibitem[\protect\citeauthoryear{He, Deng, Wang, Li, Zhang, and Wang}{He
  et~al\mbox{.}}{2020}]%
        {DBLP:conf/sigir/0001DWLZ020}
\bibfield{author}{\bibinfo{person}{Xiangnan He}, \bibinfo{person}{Kuan Deng},
  \bibinfo{person}{Xiang Wang}, \bibinfo{person}{Yan Li},
  \bibinfo{person}{Yong{-}Dong Zhang}, {and} \bibinfo{person}{Meng Wang}.}
  \bibinfo{year}{2020}\natexlab{}.
\newblock \showarticletitle{LightGCN: Simplifying and Powering Graph
  Convolution Network for Recommendation}. In
  \bibinfo{booktitle}{\emph{Proceedings of the 43rd International {ACM} {SIGIR}
  conference on research and development in Information Retrieval, {SIGIR}
  2020, Virtual Event, China, July 25-30, 2020}}. \bibinfo{publisher}{{ACM}},
  \bibinfo{pages}{639--648}.
\newblock


\bibitem[\protect\citeauthoryear{He, He, Song, Liu, Jiang, and Chua}{He
  et~al\mbox{.}}{2018}]%
        {DBLP:journals/tkde/HeHSLJC18}
\bibfield{author}{\bibinfo{person}{Xiangnan He}, \bibinfo{person}{Zhankui He},
  \bibinfo{person}{Jingkuan Song}, \bibinfo{person}{Zhenguang Liu},
  \bibinfo{person}{Yu{-}Gang Jiang}, {and} \bibinfo{person}{Tat{-}Seng Chua}.}
  \bibinfo{year}{2018}\natexlab{}.
\newblock \showarticletitle{{NAIS:} Neural Attentive Item Similarity Model for
  Recommendation}.
\newblock \bibinfo{journal}{\emph{{IEEE} Trans. Knowl. Data Eng.}}
  \bibinfo{volume}{30}, \bibinfo{number}{12} (\bibinfo{year}{2018}),
  \bibinfo{pages}{2354--2366}.
\newblock


\bibitem[\protect\citeauthoryear{He, Liao, Zhang, Nie, Hu, and Chua}{He
  et~al\mbox{.}}{2017}]%
        {DBLP:conf/www/HeLZNHC17}
\bibfield{author}{\bibinfo{person}{Xiangnan He}, \bibinfo{person}{Lizi Liao},
  \bibinfo{person}{Hanwang Zhang}, \bibinfo{person}{Liqiang Nie},
  \bibinfo{person}{Xia Hu}, {and} \bibinfo{person}{Tat{-}Seng Chua}.}
  \bibinfo{year}{2017}\natexlab{}.
\newblock \showarticletitle{Neural Collaborative Filtering}. In
  \bibinfo{booktitle}{\emph{Proceedings of the 26th International Conference on
  World Wide Web, {WWW} 2017, Perth, Australia, April 3-7, 2017}}.
  \bibinfo{publisher}{{ACM}}, \bibinfo{pages}{173--182}.
\newblock


\bibitem[\protect\citeauthoryear{Hern{\'a}ndez and Bennison}{Hern{\'a}ndez and
  Bennison}{2000}]%
        {Hernndez2000TheAA}
\bibfield{author}{\bibinfo{person}{Tony Hern{\'a}ndez} {and}
  \bibinfo{person}{David~J. Bennison}.} \bibinfo{year}{2000}\natexlab{}.
\newblock \showarticletitle{The art and science of retail location decisions}.
\newblock \bibinfo{journal}{\emph{International Journal of Retail \&
  Distribution Management}}  \bibinfo{volume}{28} (\bibinfo{year}{2000}),
  \bibinfo{pages}{357--367}.
\newblock


\bibitem[\protect\citeauthoryear{Hosmer and Lemeshow}{Hosmer and
  Lemeshow}{1991}]%
        {Hosmer1991AppliedLR}
\bibfield{author}{\bibinfo{person}{David~W. Hosmer} {and}
  \bibinfo{person}{Stanley Lemeshow}.} \bibinfo{year}{1991}\natexlab{}.
\newblock \showarticletitle{Applied Logistic Regression}.
\newblock


\bibitem[\protect\citeauthoryear{Huff}{Huff}{1966}]%
        {Huff1966APS}
\bibfield{author}{\bibinfo{person}{David~L. Huff}.}
  \bibinfo{year}{1966}\natexlab{}.
\newblock \showarticletitle{A Programmed Solution for Approximating an Optimum
  Retail Location}.
\newblock \bibinfo{journal}{\emph{Land Economics}}  \bibinfo{volume}{42}
  (\bibinfo{year}{1966}), \bibinfo{pages}{293--303}.
\newblock


\bibitem[\protect\citeauthoryear{Jaro}{Jaro}{1989}]%
        {Jaro1989AdvancesIR}
\bibfield{author}{\bibinfo{person}{Matthew~A. Jaro}.}
  \bibinfo{year}{1989}\natexlab{}.
\newblock \showarticletitle{Advances in Record-Linkage Methodology as Applied
  to Matching the 1985 Census of Tampa, Florida}.
\newblock \bibinfo{journal}{\emph{J. Amer. Statist. Assoc.}}
  \bibinfo{volume}{84} (\bibinfo{year}{1989}), \bibinfo{pages}{414--420}.
\newblock


\bibitem[\protect\citeauthoryear{Jensen}{Jensen}{2006}]%
        {Jensen2006NetworkbasedPO}
\bibfield{author}{\bibinfo{person}{Pablo Jensen}.}
  \bibinfo{year}{2006}\natexlab{}.
\newblock \showarticletitle{Network-based predictions of retail store
  commercial categories and optimal locations.}
\newblock \bibinfo{journal}{\emph{Physical review. E, Statistical, nonlinear,
  and soft matter physics}}  \bibinfo{volume}{74 3 Pt 2}
  (\bibinfo{year}{2006}), \bibinfo{pages}{035101}.
\newblock


\bibitem[\protect\citeauthoryear{Jensen}{Jensen}{2009}]%
        {DBLP:conf/ida/Jensen09}
\bibfield{author}{\bibinfo{person}{Pablo Jensen}.}
  \bibinfo{year}{2009}\natexlab{}.
\newblock \showarticletitle{Analyzing the Localization of Retail Stores with
  Complex Systems Tools}. In \bibinfo{booktitle}{\emph{Advances in Intelligent
  Data Analysis VIII, 8th International Symposium on Intelligent Data Analysis,
  {IDA} 2009, Lyon, France, August 31 - September 2, 2009. Proceedings}}
  \emph{(\bibinfo{series}{Lecture Notes in Computer Science})},
  Vol.~\bibinfo{volume}{5772}. \bibinfo{publisher}{Springer},
  \bibinfo{pages}{10--20}.
\newblock


\bibitem[\protect\citeauthoryear{Kabbur, Ning, and Karypis}{Kabbur
  et~al\mbox{.}}{2013}]%
        {DBLP:conf/kdd/KabburNK13}
\bibfield{author}{\bibinfo{person}{Santosh Kabbur}, \bibinfo{person}{Xia Ning},
  {and} \bibinfo{person}{George Karypis}.} \bibinfo{year}{2013}\natexlab{}.
\newblock \showarticletitle{{FISM:} factored item similarity models for top-N
  recommender systems}. In \bibinfo{booktitle}{\emph{The 19th {ACM} {SIGKDD}
  International Conference on Knowledge Discovery and Data Mining, {KDD} 2013,
  Chicago, IL, USA, August 11-14, 2013}}. \bibinfo{publisher}{{ACM}},
  \bibinfo{pages}{659--667}.
\newblock


\bibitem[\protect\citeauthoryear{Karamshuk, Noulas, Scellato, Nicosia, and
  Mascolo}{Karamshuk et~al\mbox{.}}{2013}]%
        {DBLP:conf/kdd/KaramshukNSNM13}
\bibfield{author}{\bibinfo{person}{Dmytro Karamshuk},
  \bibinfo{person}{Anastasios Noulas}, \bibinfo{person}{Salvatore Scellato},
  \bibinfo{person}{Vincenzo Nicosia}, {and} \bibinfo{person}{Cecilia Mascolo}.}
  \bibinfo{year}{2013}\natexlab{}.
\newblock \showarticletitle{Geo-spotting: mining online location-based services
  for optimal retail store placement}. In \bibinfo{booktitle}{\emph{The 19th
  {ACM} {SIGKDD} International Conference on Knowledge Discovery and Data
  Mining, {KDD} 2013, Chicago, IL, USA, August 11-14, 2013}}.
  \bibinfo{publisher}{{ACM}}, \bibinfo{pages}{793--801}.
\newblock


\bibitem[\protect\citeauthoryear{Kingma and Ba}{Kingma and Ba}{2015}]%
        {DBLP:journals/corr/KingmaB14}
\bibfield{author}{\bibinfo{person}{Diederik~P. Kingma} {and}
  \bibinfo{person}{Jimmy Ba}.} \bibinfo{year}{2015}\natexlab{}.
\newblock \showarticletitle{Adam: {A} Method for Stochastic Optimization}. In
  \bibinfo{booktitle}{\emph{3rd International Conference on Learning
  Representations, {ICLR} 2015, San Diego, CA, USA, May 7-9, 2015, Conference
  Track Proceedings}}.
\newblock


\bibitem[\protect\citeauthoryear{Koren, Bell, and Volinsky}{Koren
  et~al\mbox{.}}{2009}]%
        {DBLP:journals/computer/KorenBV09}
\bibfield{author}{\bibinfo{person}{Yehuda Koren}, \bibinfo{person}{Robert~M.
  Bell}, {and} \bibinfo{person}{Chris Volinsky}.}
  \bibinfo{year}{2009}\natexlab{}.
\newblock \showarticletitle{Matrix Factorization Techniques for Recommender
  Systems}.
\newblock \bibinfo{journal}{\emph{Computer}} \bibinfo{volume}{42},
  \bibinfo{number}{8} (\bibinfo{year}{2009}), \bibinfo{pages}{30--37}.
\newblock


\bibitem[\protect\citeauthoryear{Kumar and Karande}{Kumar and Karande}{2000}]%
        {Kumar2000TheEO}
\bibfield{author}{\bibinfo{person}{Prof~Vikas Kumar} {and}
  \bibinfo{person}{Kiran Karande}.} \bibinfo{year}{2000}\natexlab{}.
\newblock \showarticletitle{The Effect of Retail Store Environment on Retailer
  Performance}.
\newblock \bibinfo{journal}{\emph{Journal of Business Research}}
  \bibinfo{volume}{49} (\bibinfo{year}{2000}), \bibinfo{pages}{167--181}.
\newblock


\bibitem[\protect\citeauthoryear{Lehmann and Pan}{Lehmann and Pan}{1994}]%
        {Lehmann1994ContextEN}
\bibfield{author}{\bibinfo{person}{Donald~R. Lehmann} {and}
  \bibinfo{person}{Yigang Pan}.} \bibinfo{year}{1994}\natexlab{}.
\newblock \showarticletitle{Context Effects, New Brand Entry, and Consideration
  Sets}.
\newblock \bibinfo{journal}{\emph{Journal of Marketing Research}}
  \bibinfo{volume}{31} (\bibinfo{year}{1994}), \bibinfo{pages}{364 -- 374}.
\newblock


\bibitem[\protect\citeauthoryear{Li, Zheng, Ji, Wang, U, and Gong}{Li
  et~al\mbox{.}}{2015}]%
        {DBLP:conf/gis/LiZJWUG15}
\bibfield{author}{\bibinfo{person}{Yuhong Li}, \bibinfo{person}{Yu Zheng},
  \bibinfo{person}{Shenggong Ji}, \bibinfo{person}{Wenjun Wang},
  \bibinfo{person}{Leong~Hou U}, {and} \bibinfo{person}{Zhiguo Gong}.}
  \bibinfo{year}{2015}\natexlab{}.
\newblock \showarticletitle{Location selection for ambulance stations: a
  data-driven approach}. In \bibinfo{booktitle}{\emph{Proceedings of the 23rd
  {SIGSPATIAL} International Conference on Advances in Geographic Information
  Systems, Bellevue, WA, USA, November 3-6, 2015}}. \bibinfo{publisher}{{ACM}},
  \bibinfo{pages}{85:1--85:4}.
\newblock


\bibitem[\protect\citeauthoryear{Lian, Zhou, Zhang, Chen, Xie, and Sun}{Lian
  et~al\mbox{.}}{2018}]%
        {DBLP:conf/kdd/LianZZCXS18}
\bibfield{author}{\bibinfo{person}{Jianxun Lian}, \bibinfo{person}{Xiaohuan
  Zhou}, \bibinfo{person}{Fuzheng Zhang}, \bibinfo{person}{Zhongxia Chen},
  \bibinfo{person}{Xing Xie}, {and} \bibinfo{person}{Guangzhong Sun}.}
  \bibinfo{year}{2018}\natexlab{}.
\newblock \showarticletitle{xDeepFM: Combining Explicit and Implicit Feature
  Interactions for Recommender Systems}. In
  \bibinfo{booktitle}{\emph{Proceedings of the 24th {ACM} {SIGKDD}
  International Conference on Knowledge Discovery {\&} Data Mining, {KDD} 2018,
  London, UK, August 19-23, 2018}}. \bibinfo{publisher}{{ACM}},
  \bibinfo{pages}{1754--1763}.
\newblock


\bibitem[\protect\citeauthoryear{Liu, Wu, Zhuang, Lu, Dou, and Xiong}{Liu
  et~al\mbox{.}}{2021c}]%
        {DBLP:conf/aaai/LiuWZLD021}
\bibfield{author}{\bibinfo{person}{Hao Liu}, \bibinfo{person}{Qiyu Wu},
  \bibinfo{person}{Fuzhen Zhuang}, \bibinfo{person}{Xinjiang Lu},
  \bibinfo{person}{Dejing Dou}, {and} \bibinfo{person}{Hui Xiong}.}
  \bibinfo{year}{2021}\natexlab{c}.
\newblock \showarticletitle{Community-Aware Multi-Task Transportation Demand
  Prediction}. In \bibinfo{booktitle}{\emph{Thirty-Fifth {AAAI} Conference on
  Artificial Intelligence, {AAAI} 2021, Thirty-Third Conference on Innovative
  Applications of Artificial Intelligence, {IAAI} 2021, The Eleventh Symposium
  on Educational Advances in Artificial Intelligence, {EAAI} 2021, Virtual
  Event, February 2-9, 2021}}. \bibinfo{publisher}{{AAAI} Press},
  \bibinfo{pages}{320--327}.
\newblock


\bibitem[\protect\citeauthoryear{Liu, Ding, and Li}{Liu et~al\mbox{.}}{2021a}]%
        {DBLP:journals/corr/abs-2111-00787}
\bibfield{author}{\bibinfo{person}{Yu Liu}, \bibinfo{person}{Jingtao Ding},
  {and} \bibinfo{person}{Yong Li}.} \bibinfo{year}{2021}\natexlab{a}.
\newblock \showarticletitle{Knowledge-driven Site Selection via Urban Knowledge
  Graph}.
\newblock \bibinfo{journal}{\emph{CoRR}}  \bibinfo{volume}{abs/2111.00787}
  (\bibinfo{year}{2021}).
\newblock


\bibitem[\protect\citeauthoryear{Liu, Ding, and Li}{Liu et~al\mbox{.}}{2022}]%
        {Liu2022DevelopingKG}
\bibfield{author}{\bibinfo{person}{Yu Liu}, \bibinfo{person}{Jingtao Ding},
  {and} \bibinfo{person}{Yong Li}.} \bibinfo{year}{2022}\natexlab{}.
\newblock \showarticletitle{Developing knowledge graph based system for urban
  computing}.
\newblock \bibinfo{journal}{\emph{Proceedings of the 1st ACM SIGSPATIAL
  International Workshop on Geospatial Knowledge Graphs}}
  (\bibinfo{year}{2022}).
\newblock


\bibitem[\protect\citeauthoryear{Liu, Guo, Li, Zhang, Chen, Zhang, Liu, Yu,
  Zhang, and Yao}{Liu et~al\mbox{.}}{2019}]%
        {DBLP:journals/iotj/LiuGLZCZLYZY19}
\bibfield{author}{\bibinfo{person}{Yan Liu}, \bibinfo{person}{Bin Guo},
  \bibinfo{person}{Nuo Li}, \bibinfo{person}{Jing Zhang},
  \bibinfo{person}{Jingmin Chen}, \bibinfo{person}{Daqing Zhang},
  \bibinfo{person}{Yinxiao Liu}, \bibinfo{person}{Zhiwen Yu},
  \bibinfo{person}{Sizhe Zhang}, {and} \bibinfo{person}{Lina Yao}.}
  \bibinfo{year}{2019}\natexlab{}.
\newblock \showarticletitle{DeepStore: An Interaction-Aware Wide{\&}Deep Model
  for Store Site Recommendation With Attentional Spatial Embeddings}.
\newblock \bibinfo{journal}{\emph{{IEEE} Internet Things J.}}
  \bibinfo{volume}{6}, \bibinfo{number}{4} (\bibinfo{year}{2019}),
  \bibinfo{pages}{7319--7333}.
\newblock


\bibitem[\protect\citeauthoryear{Liu, Guo, Zhang, Zeghlache, Chen, Hu, Zhang,
  Zhou, and Yu}{Liu et~al\mbox{.}}{2021b}]%
        {DBLP:journals/tkdd/LiuGZZCHZZY21}
\bibfield{author}{\bibinfo{person}{Yan Liu}, \bibinfo{person}{Bin Guo},
  \bibinfo{person}{Daqing Zhang}, \bibinfo{person}{Djamal Zeghlache},
  \bibinfo{person}{Jingmin Chen}, \bibinfo{person}{Ke Hu},
  \bibinfo{person}{Sizhe Zhang}, \bibinfo{person}{Dan Zhou}, {and}
  \bibinfo{person}{Zhiwen Yu}.} \bibinfo{year}{2021}\natexlab{b}.
\newblock \showarticletitle{Knowledge Transfer with Weighted Adversarial
  Network for Cold-Start Store Site Recommendation}.
\newblock \bibinfo{journal}{\emph{{ACM} Trans. Knowl. Discov. Data}}
  \bibinfo{volume}{15}, \bibinfo{number}{3} (\bibinfo{year}{2021}),
  \bibinfo{pages}{47:1--47:27}.
\newblock


\bibitem[\protect\citeauthoryear{Luo, Zhou, Bao, Li, Culpepper, Ying, Liu, and
  Xiong}{Luo et~al\mbox{.}}{2020}]%
        {DBLP:conf/sigir/LuoZBLCYLX20}
\bibfield{author}{\bibinfo{person}{Hui Luo}, \bibinfo{person}{Jingbo Zhou},
  \bibinfo{person}{Zhifeng Bao}, \bibinfo{person}{Shuangli Li},
  \bibinfo{person}{J.~Shane Culpepper}, \bibinfo{person}{Haochao Ying},
  \bibinfo{person}{Hao Liu}, {and} \bibinfo{person}{Hui Xiong}.}
  \bibinfo{year}{2020}\natexlab{}.
\newblock \showarticletitle{Spatial Object Recommendation with Hints: When
  Spatial Granularity Matters}. In \bibinfo{booktitle}{\emph{Proceedings of the
  43rd International {ACM} {SIGIR} conference on research and development in
  Information Retrieval, {SIGIR} 2020, Virtual Event, China, July 25-30,
  2020}}. \bibinfo{publisher}{{ACM}}, \bibinfo{pages}{781--790}.
\newblock


\bibitem[\protect\citeauthoryear{Maesen and Lamey}{Maesen and Lamey}{2022}]%
        {Maesen2022TheIO}
\bibfield{author}{\bibinfo{person}{Stijn Maesen} {and} \bibinfo{person}{Lien
  Lamey}.} \bibinfo{year}{2022}\natexlab{}.
\newblock \showarticletitle{The Impact of Organic Specialist Store Entry on
  Category Performance at Incumbent Stores}.
\newblock \bibinfo{journal}{\emph{Journal of Marketing}}  \bibinfo{volume}{87}
  (\bibinfo{year}{2022}), \bibinfo{pages}{97 -- 113}.
\newblock


\bibitem[\protect\citeauthoryear{Marsh and Schilling}{Marsh and
  Schilling}{1994}]%
        {Marsh1994EquityMI}
\bibfield{author}{\bibinfo{person}{Michael~T. Marsh} {and}
  \bibinfo{person}{David~A. Schilling}.} \bibinfo{year}{1994}\natexlab{}.
\newblock \showarticletitle{Equity measurement in facility location analysis: A
  review and framework}.
\newblock \bibinfo{journal}{\emph{European Journal of Operational Research}}
  \bibinfo{volume}{74} (\bibinfo{year}{1994}), \bibinfo{pages}{1--17}.
\newblock


\bibitem[\protect\citeauthoryear{McFadden}{McFadden}{1977}]%
        {McFadden1977ModellingTC}
\bibfield{author}{\bibinfo{person}{Daniel McFadden}.}
  \bibinfo{year}{1977}\natexlab{}.
\newblock \showarticletitle{Modelling the Choice of Residential Location}.
\newblock \bibinfo{journal}{\emph{Transportation Research Record}}
  (\bibinfo{year}{1977}).
\newblock


\bibitem[\protect\citeauthoryear{Ning, Liu, Wang, Zeng, and Xiong}{Ning
  et~al\mbox{.}}{2023}]%
        {DBLP:journals/corr/abs-2306-11443}
\bibfield{author}{\bibinfo{person}{Yansong Ning}, \bibinfo{person}{Hao Liu},
  \bibinfo{person}{Hao Wang}, \bibinfo{person}{Zhenyu Zeng}, {and}
  \bibinfo{person}{Hui Xiong}.} \bibinfo{year}{2023}\natexlab{}.
\newblock \showarticletitle{{UUKG:} Unified Urban Knowledge Graph Dataset for
  Urban Spatiotemporal Prediction}.
\newblock \bibinfo{journal}{\emph{CoRR}}  \bibinfo{volume}{abs/2306.11443}
  (\bibinfo{year}{2023}).
\newblock


\bibitem[\protect\citeauthoryear{Pan and Lehmann}{Pan and Lehmann}{1993}]%
        {Pan1993TheIO}
\bibfield{author}{\bibinfo{person}{Yigang Pan} {and} \bibinfo{person}{Donald~R.
  Lehmann}.} \bibinfo{year}{1993}\natexlab{}.
\newblock \showarticletitle{The Influence of New Brand Entry on Subjective
  Brand Judgments}.
\newblock \bibinfo{journal}{\emph{Journal of Consumer Research}}
  \bibinfo{volume}{20} (\bibinfo{year}{1993}), \bibinfo{pages}{76--86}.
\newblock


\bibitem[\protect\citeauthoryear{Paszke, Gross, Massa, Lerer, Bradbury, Chanan,
  Killeen, Lin, Gimelshein, Antiga, Desmaison, K{\"{o}}pf, Yang, DeVito,
  Raison, Tejani, Chilamkurthy, Steiner, Fang, Bai, and Chintala}{Paszke
  et~al\mbox{.}}{2019}]%
        {DBLP:conf/nips/PaszkeGMLBCKLGA19}
\bibfield{author}{\bibinfo{person}{Adam Paszke}, \bibinfo{person}{Sam Gross},
  \bibinfo{person}{Francisco Massa}, \bibinfo{person}{Adam Lerer},
  \bibinfo{person}{James Bradbury}, \bibinfo{person}{Gregory Chanan},
  \bibinfo{person}{Trevor Killeen}, \bibinfo{person}{Zeming Lin},
  \bibinfo{person}{Natalia Gimelshein}, \bibinfo{person}{Luca Antiga},
  \bibinfo{person}{Alban Desmaison}, \bibinfo{person}{Andreas K{\"{o}}pf},
  \bibinfo{person}{Edward~Z. Yang}, \bibinfo{person}{Zachary DeVito},
  \bibinfo{person}{Martin Raison}, \bibinfo{person}{Alykhan Tejani},
  \bibinfo{person}{Sasank Chilamkurthy}, \bibinfo{person}{Benoit Steiner},
  \bibinfo{person}{Lu Fang}, \bibinfo{person}{Junjie Bai}, {and}
  \bibinfo{person}{Soumith Chintala}.} \bibinfo{year}{2019}\natexlab{}.
\newblock \showarticletitle{PyTorch: An Imperative Style, High-Performance Deep
  Learning Library}. In \bibinfo{booktitle}{\emph{Advances in Neural
  Information Processing Systems 32: Annual Conference on Neural Information
  Processing Systems 2019, NeurIPS 2019, December 8-14, 2019, Vancouver, BC,
  Canada}}. \bibinfo{pages}{8024--8035}.
\newblock


\bibitem[\protect\citeauthoryear{Pauwels and Srinivasan}{Pauwels and
  Srinivasan}{2004}]%
        {Pauwels2004WhoBF}
\bibfield{author}{\bibinfo{person}{Koen~H. Pauwels} {and}
  \bibinfo{person}{Shuba Srinivasan}.} \bibinfo{year}{2004}\natexlab{}.
\newblock \showarticletitle{Who Benefits from Store Brand Entry}.
\newblock \bibinfo{journal}{\emph{Marketing Science}}  \bibinfo{volume}{23}
  (\bibinfo{year}{2004}), \bibinfo{pages}{364--390}.
\newblock


\bibitem[\protect\citeauthoryear{Pedregosa, Varoquaux, Gramfort, Michel,
  Thirion, Grisel, Blondel, Prettenhofer, Weiss, Dubourg, Vanderplas, Passos,
  Cournapeau, Brucher, Perrot, and Duchesnay}{Pedregosa et~al\mbox{.}}{2011}]%
        {scikit-learn}
\bibfield{author}{\bibinfo{person}{F. Pedregosa}, \bibinfo{person}{G.
  Varoquaux}, \bibinfo{person}{A. Gramfort}, \bibinfo{person}{V. Michel},
  \bibinfo{person}{B. Thirion}, \bibinfo{person}{O. Grisel},
  \bibinfo{person}{M. Blondel}, \bibinfo{person}{P. Prettenhofer},
  \bibinfo{person}{R. Weiss}, \bibinfo{person}{V. Dubourg}, \bibinfo{person}{J.
  Vanderplas}, \bibinfo{person}{A. Passos}, \bibinfo{person}{D. Cournapeau},
  \bibinfo{person}{M. Brucher}, \bibinfo{person}{M. Perrot}, {and}
  \bibinfo{person}{E. Duchesnay}.} \bibinfo{year}{2011}\natexlab{}.
\newblock \showarticletitle{Scikit-learn: Machine Learning in {P}ython}.
\newblock \bibinfo{journal}{\emph{Journal of Machine Learning Research}}
  \bibinfo{volume}{12} (\bibinfo{year}{2011}), \bibinfo{pages}{2825--2830}.
\newblock


\bibitem[\protect\citeauthoryear{Porta, Latora, Wang, Rueda, Strano, Scellato,
  Cardillo, Belli, Cardenas, Cormenzana, and Latora}{Porta
  et~al\mbox{.}}{2012}]%
        {Porta2012StreetCA}
\bibfield{author}{\bibinfo{person}{Sergio Porta}, \bibinfo{person}{Vito
  Latora}, \bibinfo{person}{Fahui Wang}, \bibinfo{person}{Salvador Rueda},
  \bibinfo{person}{Emanuele Strano}, \bibinfo{person}{Salvatore Scellato},
  \bibinfo{person}{Allessio Cardillo}, \bibinfo{person}{Eugenio Belli},
  \bibinfo{person}{Francisco Cardenas}, \bibinfo{person}{Berta Cormenzana},
  {and} \bibinfo{person}{Laura Latora}.} \bibinfo{year}{2012}\natexlab{}.
\newblock \showarticletitle{Street Centrality and the Location of Economic
  Activities in Barcelona}.
\newblock \bibinfo{journal}{\emph{Urban Studies}}  \bibinfo{volume}{49}
  (\bibinfo{year}{2012}), \bibinfo{pages}{1471 -- 1488}.
\newblock


\bibitem[\protect\citeauthoryear{Porta, Strano, Iacoviello, Messora, Latora,
  Cardillo, Wang, and Scellato}{Porta et~al\mbox{.}}{2009}]%
        {Porta2009StreetCA}
\bibfield{author}{\bibinfo{person}{Sergio Porta}, \bibinfo{person}{Emanuele
  Strano}, \bibinfo{person}{Valentino Iacoviello}, \bibinfo{person}{Roberto
  Messora}, \bibinfo{person}{Vito Latora}, \bibinfo{person}{Alessio Cardillo},
  \bibinfo{person}{Fahui Wang}, {and} \bibinfo{person}{Salvatore Scellato}.}
  \bibinfo{year}{2009}\natexlab{}.
\newblock \showarticletitle{Street Centrality and Densities of Retail and
  Services in Bologna, Italy}.
\newblock \bibinfo{journal}{\emph{Environment and Planning B: Planning and
  Design}}  \bibinfo{volume}{36} (\bibinfo{year}{2009}), \bibinfo{pages}{450 --
  465}.
\newblock


\bibitem[\protect\citeauthoryear{Rathore, Ahmad, Paul, and Rho}{Rathore
  et~al\mbox{.}}{2016}]%
        {Rathore2016UrbanPA}
\bibfield{author}{\bibinfo{person}{M.~Mazhar Rathore}, \bibinfo{person}{Awais
  Ahmad}, \bibinfo{person}{Anand Paul}, {and} \bibinfo{person}{Seungmin Rho}.}
  \bibinfo{year}{2016}\natexlab{}.
\newblock \showarticletitle{Urban planning and building smart cities based on
  the Internet of Things using Big Data analytics}.
\newblock \bibinfo{journal}{\emph{Comput. Networks}}  \bibinfo{volume}{101}
  (\bibinfo{year}{2016}), \bibinfo{pages}{63--80}.
\newblock


\bibitem[\protect\citeauthoryear{Rendle, Freudenthaler, Gantner, and
  Schmidt{-}Thieme}{Rendle et~al\mbox{.}}{2009}]%
        {DBLP:conf/uai/RendleFGS09}
\bibfield{author}{\bibinfo{person}{Steffen Rendle}, \bibinfo{person}{Christoph
  Freudenthaler}, \bibinfo{person}{Zeno Gantner}, {and} \bibinfo{person}{Lars
  Schmidt{-}Thieme}.} \bibinfo{year}{2009}\natexlab{}.
\newblock \showarticletitle{{BPR:} Bayesian Personalized Ranking from Implicit
  Feedback}. In \bibinfo{booktitle}{\emph{{UAI} 2009, Proceedings of the
  Twenty-Fifth Conference on Uncertainty in Artificial Intelligence, Montreal,
  QC, Canada, June 18-21, 2009}}. \bibinfo{publisher}{{AUAI} Press},
  \bibinfo{pages}{452--461}.
\newblock


\bibitem[\protect\citeauthoryear{Revelle and Eiselt}{Revelle and
  Eiselt}{2005}]%
        {Revelle2005LocationAA}
\bibfield{author}{\bibinfo{person}{Charles~S. Revelle} {and}
  \bibinfo{person}{Horst~A. Eiselt}.} \bibinfo{year}{2005}\natexlab{}.
\newblock \showarticletitle{Location analysis: A synthesis and survey}.
\newblock \bibinfo{journal}{\emph{Eur. J. Oper. Res.}}  \bibinfo{volume}{165}
  (\bibinfo{year}{2005}), \bibinfo{pages}{1--19}.
\newblock


\bibitem[\protect\citeauthoryear{Richardson, Dominowska, and Ragno}{Richardson
  et~al\mbox{.}}{2007}]%
        {DBLP:conf/www/RichardsonDR07}
\bibfield{author}{\bibinfo{person}{Matthew Richardson}, \bibinfo{person}{Ewa
  Dominowska}, {and} \bibinfo{person}{Robert Ragno}.}
  \bibinfo{year}{2007}\natexlab{}.
\newblock \showarticletitle{Predicting clicks: estimating the click-through
  rate for new ads}. In \bibinfo{booktitle}{\emph{Proceedings of the 16th
  International Conference on World Wide Web, {WWW} 2007, Banff, Alberta,
  Canada, May 8-12, 2007}}. \bibinfo{publisher}{{ACM}},
  \bibinfo{pages}{521--530}.
\newblock


\bibitem[\protect\citeauthoryear{Sarwar, Karypis, Konstan, and Riedl}{Sarwar
  et~al\mbox{.}}{2001}]%
        {DBLP:conf/www/SarwarKKR01}
\bibfield{author}{\bibinfo{person}{Badrul~Munir Sarwar},
  \bibinfo{person}{George Karypis}, \bibinfo{person}{Joseph~A. Konstan}, {and}
  \bibinfo{person}{John Riedl}.} \bibinfo{year}{2001}\natexlab{}.
\newblock \showarticletitle{Item-based collaborative filtering recommendation
  algorithms}. In \bibinfo{booktitle}{\emph{Proceedings of the Tenth
  International World Wide Web Conference, {WWW} 10, Hong Kong, China, May 1-5,
  2001}}. \bibinfo{publisher}{{ACM}}, \bibinfo{pages}{285--295}.
\newblock


\bibitem[\protect\citeauthoryear{Seenivasan and Talukdar}{Seenivasan and
  Talukdar}{2016}]%
        {Seenivasan2016CompetitiveEO}
\bibfield{author}{\bibinfo{person}{Satheesh Seenivasan} {and}
  \bibinfo{person}{Debabrata Talukdar}.} \bibinfo{year}{2016}\natexlab{}.
\newblock \showarticletitle{Competitive Effects of Wal-Mart Supercenter Entry:
  Moderating Roles of Category and Brand Characteristics}.
\newblock \bibinfo{journal}{\emph{Journal of Retailing}}  \bibinfo{volume}{92}
  (\bibinfo{year}{2016}), \bibinfo{pages}{218--225}.
\newblock


\bibitem[\protect\citeauthoryear{Shao, Han, Sun, Xiao, lei Zhang, and
  xu~Zhao}{Shao et~al\mbox{.}}{2020}]%
        {Shao2020ARO}
\bibfield{author}{\bibinfo{person}{Meng Shao}, \bibinfo{person}{Zhi Han},
  \bibinfo{person}{Jin~Wei Sun}, \bibinfo{person}{Chengsi Xiao},
  \bibinfo{person}{Shu lei Zhang}, {and} \bibinfo{person}{Yuan xu Zhao}.}
  \bibinfo{year}{2020}\natexlab{}.
\newblock \showarticletitle{A review of multi-criteria decision making
  applications for renewable energy site selection}.
\newblock \bibinfo{journal}{\emph{Renewable Energy}}  \bibinfo{volume}{157}
  (\bibinfo{year}{2020}), \bibinfo{pages}{377--403}.
\newblock


\bibitem[\protect\citeauthoryear{van~den Berg, Kipf, and Welling}{van~den Berg
  et~al\mbox{.}}{2017}]%
        {DBLP:journals/corr/BergKW17}
\bibfield{author}{\bibinfo{person}{Rianne van~den Berg},
  \bibinfo{person}{Thomas~N. Kipf}, {and} \bibinfo{person}{Max Welling}.}
  \bibinfo{year}{2017}\natexlab{}.
\newblock \showarticletitle{Graph Convolutional Matrix Completion}.
\newblock \bibinfo{journal}{\emph{CoRR}}  \bibinfo{volume}{abs/1706.02263}
  (\bibinfo{year}{2017}).
\newblock


\bibitem[\protect\citeauthoryear{Wang, He, Wang, Feng, and Chua}{Wang
  et~al\mbox{.}}{2019}]%
        {DBLP:conf/sigir/Wang0WFC19}
\bibfield{author}{\bibinfo{person}{Xiang Wang}, \bibinfo{person}{Xiangnan He},
  \bibinfo{person}{Meng Wang}, \bibinfo{person}{Fuli Feng}, {and}
  \bibinfo{person}{Tat{-}Seng Chua}.} \bibinfo{year}{2019}\natexlab{}.
\newblock \showarticletitle{Neural Graph Collaborative Filtering}. In
  \bibinfo{booktitle}{\emph{Proceedings of the 42nd International {ACM} {SIGIR}
  Conference on Research and Development in Information Retrieval, {SIGIR}
  2019, Paris, France, July 21-25, 2019}}. \bibinfo{publisher}{{ACM}},
  \bibinfo{pages}{165--174}.
\newblock


\bibitem[\protect\citeauthoryear{Wu, Xie, Xu, and Li}{Wu et~al\mbox{.}}{2017}]%
        {Wu2017ADF}
\bibfield{author}{\bibinfo{person}{Yunna Wu}, \bibinfo{person}{Chao Xie},
  \bibinfo{person}{Chuanbo Xu}, {and} \bibinfo{person}{Fang Li}.}
  \bibinfo{year}{2017}\natexlab{}.
\newblock \showarticletitle{A Decision Framework for Electric Vehicle Charging
  Station Site Selection for Residential Communities under an Intuitionistic
  Fuzzy Environment: A Case of Beijing}.
\newblock \bibinfo{journal}{\emph{Energies}}  \bibinfo{volume}{10}
  (\bibinfo{year}{2017}), \bibinfo{pages}{1270}.
\newblock


\bibitem[\protect\citeauthoryear{Wu, Yang, Zhang, Chen, and Wang}{Wu
  et~al\mbox{.}}{2016}]%
        {Wu2016OptimalSS}
\bibfield{author}{\bibinfo{person}{Yunna Wu}, \bibinfo{person}{Meng Yang},
  \bibinfo{person}{Haobo Zhang}, \bibinfo{person}{Kaifeng Chen}, {and}
  \bibinfo{person}{Yang Wang}.} \bibinfo{year}{2016}\natexlab{}.
\newblock \showarticletitle{Optimal Site Selection of Electric Vehicle Charging
  Stations Based on a Cloud Model and the PROMETHEE Method}.
\newblock \bibinfo{journal}{\emph{Energies}}  \bibinfo{volume}{9}
  (\bibinfo{year}{2016}), \bibinfo{pages}{1--20}.
\newblock


\bibitem[\protect\citeauthoryear{Xu, Wang, Wu, Zhou, Li, and Wu}{Xu
  et~al\mbox{.}}{2016}]%
        {DBLP:conf/gis/XuWWZLW16}
\bibfield{author}{\bibinfo{person}{Mengwen Xu}, \bibinfo{person}{Tianyi Wang},
  \bibinfo{person}{Zhengwei Wu}, \bibinfo{person}{Jingbo Zhou},
  \bibinfo{person}{Jian Li}, {and} \bibinfo{person}{Haishan Wu}.}
  \bibinfo{year}{2016}\natexlab{}.
\newblock \showarticletitle{Demand driven store site selection via multiple
  spatial-temporal data}. In \bibinfo{booktitle}{\emph{Proceedings of the 24th
  {ACM} {SIGSPATIAL} International Conference on Advances in Geographic
  Information Systems, {GIS} 2016, Burlingame, California, USA, October 31 -
  November 3, 2016}}. \bibinfo{publisher}{{ACM}}, \bibinfo{pages}{40:1--40:10}.
\newblock


\bibitem[\protect\citeauthoryear{Yan, Wang, Yang, Guo, He, and Zhang}{Yan
  et~al\mbox{.}}{2022}]%
        {DBLP:conf/icde/YanWYGHZ22}
\bibfield{author}{\bibinfo{person}{Hua Yan}, \bibinfo{person}{Shuai Wang},
  \bibinfo{person}{Yu Yang}, \bibinfo{person}{Baoshen Guo},
  \bibinfo{person}{Tian He}, {and} \bibinfo{person}{Desheng Zhang}.}
  \bibinfo{year}{2022}\natexlab{}.
\newblock \showarticletitle{{\textdollar}O{\^{}}\{2\}{\textdollar}-SiteRec:
  Store Site Recommendation under the {O2O} Model via Multi-graph Attention
  Networks}. In \bibinfo{booktitle}{\emph{38th {IEEE} International Conference
  on Data Engineering, {ICDE} 2022, Kuala Lumpur, Malaysia, May 9-12, 2022}}.
  \bibinfo{publisher}{{IEEE}}, \bibinfo{pages}{525--538}.
\newblock


\bibitem[\protect\citeauthoryear{Ye, Sun, Du, Fu, and Xiong}{Ye
  et~al\mbox{.}}{2021}]%
        {DBLP:conf/aaai/YeSDF021}
\bibfield{author}{\bibinfo{person}{Junchen Ye}, \bibinfo{person}{Leilei Sun},
  \bibinfo{person}{Bowen Du}, \bibinfo{person}{Yanjie Fu}, {and}
  \bibinfo{person}{Hui Xiong}.} \bibinfo{year}{2021}\natexlab{}.
\newblock \showarticletitle{Coupled Layer-wise Graph Convolution for
  Transportation Demand Prediction}. In \bibinfo{booktitle}{\emph{Thirty-Fifth
  {AAAI} Conference on Artificial Intelligence, {AAAI} 2021, Thirty-Third
  Conference on Innovative Applications of Artificial Intelligence, {IAAI}
  2021, The Eleventh Symposium on Educational Advances in Artificial
  Intelligence, {EAAI} 2021, Virtual Event, February 2-9, 2021}}.
  \bibinfo{publisher}{{AAAI} Press}, \bibinfo{pages}{4617--4625}.
\newblock


\bibitem[\protect\citeauthoryear{Yu, Ai, and Shao}{Yu et~al\mbox{.}}{2015}]%
        {Yu2015TheAA}
\bibfield{author}{\bibinfo{person}{Wenhao Yu}, \bibinfo{person}{Tinghua Ai},
  {and} \bibinfo{person}{Shiwei Shao}.} \bibinfo{year}{2015}\natexlab{}.
\newblock \showarticletitle{The analysis and delimitation of Central Business
  District using network kernel density estimation}.
\newblock \bibinfo{journal}{\emph{Journal of Transport Geography}}
  \bibinfo{volume}{45} (\bibinfo{year}{2015}), \bibinfo{pages}{32--47}.
\newblock


\bibitem[\protect\citeauthoryear{Yu, Tian, Wang, Guo, and Mei}{Yu
  et~al\mbox{.}}{2016}]%
        {DBLP:journals/tkdd/YuTWGM16}
\bibfield{author}{\bibinfo{person}{Zhiwen Yu}, \bibinfo{person}{Miao Tian},
  \bibinfo{person}{Zhu Wang}, \bibinfo{person}{Bin Guo}, {and}
  \bibinfo{person}{Tao Mei}.} \bibinfo{year}{2016}\natexlab{}.
\newblock \showarticletitle{Shop-Type Recommendation Leveraging the Data from
  Social Media and Location-Based Services}.
\newblock \bibinfo{journal}{\emph{{ACM} Trans. Knowl. Discov. Data}}
  \bibinfo{volume}{11}, \bibinfo{number}{1} (\bibinfo{year}{2016}),
  \bibinfo{pages}{1:1--1:21}.
\newblock


\bibitem[\protect\citeauthoryear{Zhang, Zhong, Yuan, and Xiong}{Zhang
  et~al\mbox{.}}{2021b}]%
        {DBLP:conf/kdd/ZhangZYX21}
\bibfield{author}{\bibinfo{person}{Shengming Zhang}, \bibinfo{person}{Hao
  Zhong}, \bibinfo{person}{Zixuan Yuan}, {and} \bibinfo{person}{Hui Xiong}.}
  \bibinfo{year}{2021}\natexlab{b}.
\newblock \showarticletitle{Scalable Heterogeneous Graph Neural Networks for
  Predicting High-potential Early-stage Startups}. In
  \bibinfo{booktitle}{\emph{{KDD} '21: The 27th {ACM} {SIGKDD} Conference on
  Knowledge Discovery and Data Mining, Virtual Event, Singapore, August 14-18,
  2021}}. \bibinfo{publisher}{{ACM}}, \bibinfo{pages}{2202--2211}.
\newblock


\bibitem[\protect\citeauthoryear{Zhang, Zhu, and Gou}{Zhang
  et~al\mbox{.}}{2017}]%
        {Zhang2017DemandFA}
\bibfield{author}{\bibinfo{person}{Ting Zhang}, \bibinfo{person}{Xiaowei Zhu},
  {and} \bibinfo{person}{Qinglong Gou}.} \bibinfo{year}{2017}\natexlab{}.
\newblock \showarticletitle{Demand Forecasting and Pricing Decision with the
  Entry of Store Brand under Various Information Sharing Scenarios}.
\newblock \bibinfo{journal}{\emph{Asia Pac. J. Oper. Res.}}
  \bibinfo{volume}{34} (\bibinfo{year}{2017}),
  \bibinfo{pages}{1740018:1--1740018:26}.
\newblock


\bibitem[\protect\citeauthoryear{Zhang, Liu, Han, Ge, and Xiong}{Zhang
  et~al\mbox{.}}{2022}]%
        {DBLP:conf/kdd/00030HG022}
\bibfield{author}{\bibinfo{person}{Weijia Zhang}, \bibinfo{person}{Hao Liu},
  \bibinfo{person}{Jindong Han}, \bibinfo{person}{Yong Ge}, {and}
  \bibinfo{person}{Hui Xiong}.} \bibinfo{year}{2022}\natexlab{}.
\newblock \showarticletitle{Multi-Agent Graph Convolutional Reinforcement
  Learning for Dynamic Electric Vehicle Charging Pricing}. In
  \bibinfo{booktitle}{\emph{{KDD} '22: The 28th {ACM} {SIGKDD} Conference on
  Knowledge Discovery and Data Mining, Washington, DC, USA, August 14 - 18,
  2022}}. \bibinfo{publisher}{{ACM}}, \bibinfo{pages}{2471--2481}.
\newblock


\bibitem[\protect\citeauthoryear{Zhang, Liu, Wang, Xu, Xin, Dou, and
  Xiong}{Zhang et~al\mbox{.}}{2021a}]%
        {DBLP:conf/www/ZhangLWXXDX21}
\bibfield{author}{\bibinfo{person}{Weijia Zhang}, \bibinfo{person}{Hao Liu},
  \bibinfo{person}{Fan Wang}, \bibinfo{person}{Tong Xu},
  \bibinfo{person}{Haoran Xin}, \bibinfo{person}{Dejing Dou}, {and}
  \bibinfo{person}{Hui Xiong}.} \bibinfo{year}{2021}\natexlab{a}.
\newblock \showarticletitle{Intelligent Electric Vehicle Charging
  Recommendation Based on Multi-Agent Reinforcement Learning}. In
  \bibinfo{booktitle}{\emph{{WWW} '21: The Web Conference 2021, Virtual Event /
  Ljubljana, Slovenia, April 19-23, 2021}}. \bibinfo{publisher}{{ACM} /
  {IW3C2}}, \bibinfo{pages}{1856--1867}.
\newblock


\bibitem[\protect\citeauthoryear{Zhou, Yin, Zhang, Trajcevski, Zhong, and
  Wu}{Zhou et~al\mbox{.}}{2019}]%
        {DBLP:conf/www/0002YZTZW19}
\bibfield{author}{\bibinfo{person}{Fan Zhou}, \bibinfo{person}{Ruiyang Yin},
  \bibinfo{person}{Kunpeng Zhang}, \bibinfo{person}{Goce Trajcevski},
  \bibinfo{person}{Ting Zhong}, {and} \bibinfo{person}{Jin Wu}.}
  \bibinfo{year}{2019}\natexlab{}.
\newblock \showarticletitle{Adversarial Point-of-Interest Recommendation}. In
  \bibinfo{booktitle}{\emph{The World Wide Web Conference, {WWW} 2019, San
  Francisco, CA, USA, May 13-17, 2019}}. \bibinfo{publisher}{{ACM}},
  \bibinfo{pages}{3462--34618}.
\newblock


\bibitem[\protect\citeauthoryear{Zobel and Dart}{Zobel and Dart}{1996}]%
        {DBLP:conf/sigir/ZobelD96}
\bibfield{author}{\bibinfo{person}{Justin Zobel} {and}
  \bibinfo{person}{Philip~W. Dart}.} \bibinfo{year}{1996}\natexlab{}.
\newblock \showarticletitle{Phonetic String Matching: Lessons from Information
  Retrieval}. In \bibinfo{booktitle}{\emph{Proceedings of the 19th Annual
  International {ACM} {SIGIR} Conference on Research and Development in
  Information Retrieval, SIGIR'96, August 18-22, 1996, Zurich, Switzerland
  (Special Issue of the {SIGIR} Forum)}}. \bibinfo{publisher}{{ACM}},
  \bibinfo{pages}{166--172}.
\newblock


\bibitem[\protect\citeauthoryear{Şahin, Ocak, and Top}{Şahin
  et~al\mbox{.}}{2019}]%
        {ahin2019AnalyticHP}
\bibfield{author}{\bibinfo{person}{Tezcan~Kaşmer Şahin},
  \bibinfo{person}{Saffet Ocak}, {and} \bibinfo{person}{Mehmet Top}.}
  \bibinfo{year}{2019}\natexlab{}.
\newblock \showarticletitle{Analytic hierarchy process for hospital site
  selection}.
\newblock \bibinfo{journal}{\emph{Health Policy and Technology}}
  (\bibinfo{year}{2019}).
\newblock


\end{thebibliography}

\end{document}